\documentclass[11pt,a4paper]{article}
\usepackage{amsmath,amssymb}
\usepackage[pdftex]{graphicx}
\usepackage{float}
\usepackage{color}
\usepackage[utf8]{inputenc}

\makeatletter
\def\appendix{
\def\theequation{\Alph{section}\arabic{equation}}
\setcounter{equation}{0}
\def\thesection{\Alph{section}}
\@addtoreset{equation}{section}
\setcounter{section}{0}
\def\@seccntformat##1{
\@nameuse{prefix@##1}
\@nameuse{the##1}
\@nameuse{postfix@##1}\quad}
\def\prefix@section{Appendix~}
} 
\makeatother

\allowdisplaybreaks[2] % permit pagebreak in the gather or the align enviroment

\def\bp{{\boldsymbol p}}

\newcommand{\ltsim}{\protect\raisebox{-0.5ex}{$\:\stackrel{\textstyle <}{\sim}\:$}}

\def\ct{\cos \theta}
\def\tt{\tilde{\tau}}

\newcommand \beq{\begin{eqnarray}}
\newcommand \eeq{\end{eqnarray}}

\begin{document}

\date{}

\title{\bf Angular mode expansion of the Boltzmann equation in the small-angle approximation}

\author{Jean-Paul Blaizot${}^{~a}$, Naoto Tanji${}^{~b}$}
\maketitle

\begin{center}
 \begin{itemize}
  \item[{\bf a.}] Institut de Physique Th\'eorique, Universit\'e Paris-Saclay\\
 CEA, CNRS, F-91191 Gif-sur-Yvette, France
  \item[{\bf b.}] European Centre for Theoretical Studies in Nuclear Physics\\ and Related Areas (ECT*
) and Fondazione Bruno Kessler\\
Villa Tambosi, Strada delle Tabarelle 286, I-38123 Villazzano, Italy
 \end{itemize}
\end{center}
\vglue 1cm

\begin{abstract}
We  use an expansion in angular mode functions in order to solve the Boltzmann equation for a gluon plasma undergoing longitudinal expansion. By comparing with the exact solution obtained numerically by other means we show that the expansion in mode functions converges rapidly for all cases of practical interest, and represents a substantial gain in numerical effort as compared to more standard methods. We contrast the cases of a non expanding plasma and of longitudinally expanding plasmas, and follow in both cases the evolutions towards thermalization. In the latter case, we observe that, although the spherical mode function appears to be well reproduced after some time by a local equilibrium distribution function depending on slowly varying temperature and chemical potential, thereby suggesting thermalization of the system, the longitudinal and transverse pressures take more time to equilibrate. This is because the expansion hinders the relaxation of the first angular mode function. This feature was also observed in a simpler context where the Boltzmann equation is solved in terms of special moments within the relaxation time approximation, and attributed there to the particular coupling between the first two moments of the distribution function. The present analysis confirms this observation in a more realistic setting.

\end{abstract}

%%%%%%%%%%%%%%%%%%%%%%%%%%%%%%%%%%%%%%%%%%%%%%%%%%%%%%%%%%%%%%%%%%%%%%%%%%%%%%%%%%%%%%%
\section{Introduction}

This paper concerns the kinetic description of expanding quark-gluon plasmas, an area where there has been a lot of activity over the last years. The goals of these calculations is to understand the behavior of matter produced in ultra-relativistic heavy ion collisions, in particular the emergence of  hydrodynamics at late time. As representative of some recent works on the subject, let us mention those which concern the solution of the  Boltzmann equation with elastic and inelastic processes (see e.g.  \cite{Xu:2004mz,Scardina:2014gxa,Kurkela:2015qoa}), or the recent work focusing on very small coupling, of very high occupancies where simple scaling laws emerge \cite{Mazeliauskas:2018yef}. 
Aside from such elaborate simulations which aim at a realistic description of physical systems, while  incorporating detailed QCD contributions to the collision kernel, there is a need to simplify the theoretical description, and get under good theoretical control the various important physical phenomena that are involved, and their associated time scales. The present work is an effort in this direction.   

We shall consider a simple setting, that of an expanding plasma composed of gluons only, and with Bjorken symmetry \cite{Bjorken:1982qr}. Our main concern will be to understand how the momentum distribution gets isotropic as a result of the collisions, and this in spite of the strong longitudinal expansion. One of our goal here is to extend the analysis carried out in \cite{Blaizot:2017ucy,Blaizot:2019}, where it has been pointed out that, within kinetic theory,  the competition between free streaming and collisions could be understood in terms of equations for simple moments of the distribution function. These  moments, which we called ${\cal L}$-moments,  were introduced in \cite{Blaizot:2017lht}. Their knowledge is not sufficient to fully reconstruct the distribution functions, but they capture the essential features of the angular dynamics.   In the case where the collision kernel of the kinetic equation is given by a relaxation time approximation, the equations for the  ${\cal L}$-moments form an infinite hierarchy that can be solved by truncation \cite{Blaizot:2017ucy}.  Such a procedure appears to converge quickly, and  even the lowest order truncation that takes into account only the first non trivial distortion of the momentum distribution suffices to get an accurate description. We believe that this particular structure is robust, and part of our purpose in this paper is to show that this is indeed the case, by examining a more realistic collision kernel, that of elastic QCD interactions within the small scattering angle approximation \cite{Mueller:1999pi,Bjoraker:2000cf,Blaizot:2013lga}. 

When such a collision kernel is taken into account, the equations for the ${\cal L}$-moments do not close anymore, as new types of moments appear. This is because the ${\cal L}$-moments do not fully take into account the radial momentum dependence (only the angular part is treated exactly, the radial  momentum being treated in an average way). There exists of course a number of methods based on more complete moments that could be envisaged (see e.g.  \cite{Denicol:2012cn,Behtash:2017wqg}).  However, we want to retain the simplicity of the angular structure that emerged in our previous analysis, and explore another approach in which the distribution function is expanded in angular mode functions $f_n(p)$. The basic assumption is  that only a few terms in this expansion  are needed, as was the case in the expansion in ${\cal L}$-moments. We shall see that this is indeed the case, which brings a significant simplification in the solution of the Boltzmann equation, since, for the case considered, it reduces a two-dimensional equation  to a finite (small) set of one-dimensional equations.  

The outline of this paper is as follows. In the next section, we describe the angular mode expansion that we shall be using, and present the resulting form of the kinetic equation which turns into an infinite hierarchy of coupled equations. In the following section, we discuss the initial conditions that are used in the numerical calculations, and we give an estimate of the relaxation time  scale, or equivalently of the collision rate. The qualitative features of the evolution of the system in the expanding case depend on the ratio of this collision rate to the expansion rate.  We then present the results of a set of numerical calculations. First we consider the non expanding case, where we can analyze in details the isotropization of the momentum distribution and the approach to equilibrium. Then we turn to the case of a longitudinally expanding system, where  the approach to equilibrium is studied  as a function of the ratio between the collision rate and the expansion rate. Some technical material is presented in two Appendices.

%%%%%%%%%%%%%%%%%%%%%%%%%%%%%%%%%%%%%%%%%%%%%%%%%%%%%%%%%%%%%%%%%%%%%%%%%%%%%%%%%%%%%%%
\section{Angular mode expansion} \label{sec:expansion}
The Boltzmann equation for  a longitudinally expanding system takes the form  
\begin{equation}
\left( \frac{\partial}{\partial \tau} -\frac{p_z}{\tau} \frac{\partial}{\partial p_z} \right) f(\tau ,\bp ) = -C[f] \, ,
\label{Boltz0}
\end{equation}
where $f(\tau, \bp)$ denotes the single particle distribution function and $C[f]$  is the collision kernel.  We have assumed here that the system is boost-invariant in the longitudinal $z$-direction, and uniform in the transverse $(x,y)$-plane. The distribution function $f$ depends then only on the proper time $\tau=\sqrt{t^2-z^2}$ and on the momentum $\bp =(p_x,p_y,p_z)$. In writing Eq.~(\ref{Boltz0}), we have also exploited boost invariance, and Eq.~(\ref{Boltz0}) holds in the central slice around $z=0$.

We further assume that the momentum distribution function is isotropic in the $\bp_\perp =(p_x,p_y)$-plane. In this case, we can express it as a function of
\begin{equation}
p\equiv |\bp| \hspace{10pt} \text{and} \hspace{10pt} \ct \equiv \frac{p_z}{p} \, .
\end{equation}

In Eq.~(\ref{Boltz0}),  the drift term $-\frac{p_z}{\tau} \frac{\partial}{\partial p_z} f$ is responsible for the effects of the longitudinal expansion which tends to make the momentum distribution anisotropic, while the collision term tends to make it isotropic. Our primary interest in this paper is to study the competition of these two effects, and follow the evolution towards   isotropization and   hydrodynamic behavior.
To do so, we generalize the approach initiated in Refs.~\cite{Blaizot:2017lht,Blaizot:2017ucy} which is based on the following moments of the distribution function
\begin{equation}
\mathcal{L}_n (\tau) = \int\! \frac{d^3 p}{(2\pi)^3} \, p f(\tau, \bp ) P_{2n} (\ct) \, ,
\label{def_momL}
\end{equation}
where $P_{2n} (z)$ is the Legendre polynomial of order $2n$. As we assume the distribution function to be a symmetric function of $p_z$, only polynomials of even orders are necessary. 
The zeroth order moment is nothing but the energy density, $\mathcal{L}_0 =\mathcal{E}$, while the first moment gives the difference between the longitudinal and the transverse pressures, $\mathcal{L}_1 =\mathcal{P}_L -\mathcal{P}_T$. 
%Collisions tend to damp the moments of the momentum distribution, in particular $\mathcal{L}_1$, thereby equilibrating the two pressures.  

If one employs the relaxation time approximation $C[f]=(f-f_{\rm eq})/\tau_R$ for the collision term  one can derive a set of coupled equations for the moments ${\cal L}_n$, which can be solved by truncations \cite{Blaizot:2017ucy}. However, for a more realistic collision kernel, moments other than the $\mathcal{L}_n$'s appear,  which complicates the analysis. This is the case in particular when one uses the collision term  describing gluon elastic collisions in the small-angle approximation \cite{Mueller:1999pi,Blaizot:2013lga}. As shown in Appendix~\ref{sec:moment}, the equations for the moments $\mathcal{L}_n$ are then coupled to other kinds of moments, and no obvious and simple scheme emerges that would allow us to close the resulting system of equations. 

Of course, the Boltzmann equation \eqref{Boltz0} can be solved directly,  with the elastic collision term treated within the small-angle approximation (see e.g. \cite{Mueller:1999pi,Tanji:2017suk}), or beyond \cite{Scardina:2014gxa,Epelbaum:2015vxa}, as well as with  both elastic and inelastic processes included in the collision kernel (see e.g. Refs.~\cite{Xu:2004mz,Kurkela:2015qoa,Keegan:2015avk}). However these approaches are very demanding numerically. 
In this paper, we then attempt an intermediate approach between the moment method of Ref.~\cite{Blaizot:2017ucy} and the direct analysis of the full Boltzmann equation. We make a mode expansion with respect to the angular variable $\ct =p_z /p$ while the dependence on the absolute value of the 3-dimensional momentum $p=|\bp|$ is treated fully. As we shall see, since in most cases of practical interest a few mode functions are sufficient to capture the main physics, this represents a substantial gain: a two-dimensional kinetic equation is replaced by a finite (small) set of coupled one-dimensional equations.\footnote{To appreciate the numerical gain, note that in Ref.~\cite{Tanji:2017suk}, the angular variable is discretized on a grid with 256 points, whereas only a few modes functions are needed to get an accurate solution.} Aside from this technical aspect, the mode functions may represent an interesting set of degrees of freedom that provide useful insight into the dynamics of the expanding system. 

We consider the following expansion of the momentum distribution function,
\begin{equation}
f(\tau, p ,\ct) = \sum_{n=0}^\infty f_n (\tau,p) P_{2n} (\ct) \, .
\label{def_expansion}
\end{equation}
The mode function $f_n (\tau,p)$ can be extracted from the full distribution function by
\begin{equation}
f_n (\tau,p) = \frac{4n+1}{2} \int_{-1}^1 dx \, f(\tau, p, x) P_{2n} (x) \, .
\label{def_mode}
\end{equation}
In terms of the mode functions, the moments \eqref{def_momL} are given by 
\begin{equation}\label{Lnfn}
\mathcal{L}_n (\tau) = \frac{2}{4n+1} \frac{1}{4\pi^2} \int_0^\infty \! dp \, p^{3} f_n (\tau,p) \, .
\end{equation}
The particle number density ${\cal N}$ and the energy density $\mathcal{E}$ depend only on the zeroth mode function $f_0 (\tau,p)$, and are given respectively by 
\begin{align}
{\cal N} = \frac{\nu_\text{g}}{2\pi^2} \int_0^\infty \! dp \, p^2 f_0 (\tau,p) \, ,
\label{num}
\end{align}
and 
\begin{equation}
\mathcal{E} = \frac{\nu_\text{g}}{2\pi^2} \int_0^\infty \! dp \, p^3 f_0 (\tau,p) \, ,
\label{ene}
\end{equation}
where $\nu_\text{g} =2(N_c^2-1)=16$ denotes the number of internal degrees of freedom of the gluons. 
The longitudinal and the transverse pressures depend on both the zeroth and first mode functions. In terms of the moments ${\cal L}_0$ and ${\cal L}_1$ they read (see Eq.~(\ref{Lnfn}))
\begin{align}
\mathcal{P}_L = \nu_\text{g} \frac{\mathcal{L}_0 +2\mathcal{L}_1}{3} \, , \hspace{10pt}
\mathcal{P}_T = \nu_\text{g} \frac{\mathcal{L}_0 -\mathcal{L}_1}{3} \, .  
\end{align}

In the small-angle approximation, the collision term for the gluon-gluon elastic scattering takes the form  \cite{Mueller:1999pi,Blaizot:2013lga}
\begin{equation}
C[f(\tau ,\bp )] = -4\pi N_c^2 \alpha_s^2 \,  l_{\rm Cb}\, \nabla_\bp \cdot \left[ I_a \nabla_\bp f(\tau ,\bp ) +I_b \frac{\bp}{p} f(\tau ,\bp ) (1+f(\tau ,\bp )) \right] \, ,
\label{col_sa0}
\end{equation}
where $\alpha_s$ is the strong coupling constant and $l_{\rm Cb}$ is the Coulomb logarithm, treated here as a constant. The integrals $I_a$ and $I_b$ are defined by
\begin{equation}
I_a (\tau)  = \int\! \frac{d^3 p}{(2\pi)^3} \, f(\tau, \bp ) \left[  1+f(\tau, \bp ) \right]
\label{Ia}
\end{equation}
and
\begin{equation}
I_b (\tau)  = 2\int\! \frac{d^3 p}{(2\pi)^3} \, \frac{1}{p} f(\tau, \bp ) \, .
\label{Ib}
\end{equation}
Since the integrand of $I_b$ is isotropic in $\bp$ and linear in $f$,  $I_b$ can be expressed  in terms of  $f_0$ alone:
\begin{equation}
I_b (\tau) = \frac{1}{\pi^2} \int_0^\infty \! dp \, p f_0 (\tau,p) \, . 
\label{Ib_mode}
\end{equation}
In contrast, all the mode functions contribute to the integral $I_a$ as it is nonlinear in $f$,
\begin{equation}
I_a (\tau )  = \frac{1}{2\pi^2} \int_0^\infty \! dp \, p^2 \left\{ f_0 (\tau,p) 
+\sum_{n=0}^\infty \frac{1}{4n+1} \left[ f_n (\tau,p)\right]^2 \right\} \, . 
\label{Ia_mode}
\end{equation}

By applying the mode projection operator $\frac{4n+1}{2} \int_{-1}^1 d(\ct) \, P_{2n} (\ct)$ to the Boltzmann equation \eqref{Boltz0}, we can derive the kinetic equations for the mode function $f_n (\tau ,p)$  
\begin{align}
\frac{\partial}{\partial \tt} f_n (p) 
&= \frac{1}{\tt} p \frac{\partial}{\partial p} \left[ a_n f_n (p) +b_n f_{n-1} (p) +c_n f_{n+1}(p)\right] 
+\frac{1}{\tt} \left[ \tilde{a}_n f_n (p) +\tilde{b}_n f_{n-1} (p) +\tilde{c}_n f_{n+1} (p) \right] \notag \\
&\hspace{10pt} -C_n [f] \, ,
%:
%:
\label{Boltz_n}
\end{align}
where the collision term reads
\begin{align}
C_n [f(p)] 
&= -I_a  \frac{1}{p^2} \left[\frac{\partial}{\partial p} p^2 \frac{\partial}{\partial p} -2n(2n+1) \right] f_n (p) \notag \\
&\hspace{10pt}
-I_b \frac{1}{p^2} \frac{\partial}{\partial p} p^2 \left[ f_n (p)
+\frac{4n+1}{2} \sum_{m,l=0}^\infty A_{n,m,l} f_m(p) f_l(p) \right]  \, ,
\label{coll_n}
\end{align}
and we have  rescaled the time variable, setting
\begin{equation}\label{tildetau}
\tt = 4\pi N_c^2 \alpha_s^2 \, l_{\rm Cb} \, \tau \, .
\end{equation}
A further rescaling will be performed later in order to measure time in units of a suitably defined relaxation time (see Eq.~(\ref{tauR0}) below). 
Here and in the following, the time argument of $f$ is often omitted to alleviate the notation. 
The numerical coefficients that appear in the drift terms are given by
\beq\label{coeff_first}
a_n = \frac{8n^2+4n-1}{(4n-1)(4n+3)} ,\quad 
b_n = \frac{(2n-1)2n}{(4n-3)(4n-1)} ,\quad
c_n = \frac{(2n+1)(2n+2)}{(4n+3)(4n+5)} ,
\eeq
and
\beq
\tilde{a}_n = \frac{2n(2n+1)}{(4n-1)(4n+3)} ,\quad 
\tilde{b}_n = -\frac{(2n-2)(2n-1)2n}{(4n-3)(4n-1)} ,\quad
\tilde{c}_n = \frac{(2n+1)(2n+2)(2n+3)}{(4n+3)(4n+5)} \, . \label{coeff_last}
\eeq
The coefficient $A_{n,m,l}$ in the collision term is defined by
\begin{align}
A_{n,m,l}
= \int_{-1}^1 dx \, P_{2n} (x) P_{2m} (x) P_{2l} (x) \, .
\end{align}
The derivation of these equations and the explicit form of $A_{n,m,l}$ can be found in Appendix~\ref{sec:derivation}.\\

The kinetic equations for the mode functions, Eq.~\eqref{Boltz_n}, are coupled to each other through the drift terms ($\propto 1/\tau$),  the nonlinear term  in the collision kernel,  as well as the integral $I_a$. They constitute an infinite hierarchy of equations, and in order to solve them in practice we need to implement some truncation. 
In Ref.~\cite{Blaizot:2017ucy}, it is shown that such truncation works efficiently already at a low order in the moment method. In Sec.~\ref{sec:results}, we will demonstrate that a low order truncation works successfully also for the angular mode expansion.  The truncation that we shall consider amounts to solving the equations for the first $N+1$ mode functions \eqref{Boltz_n}, i.e.,  for $n=0,\cdots ,N$ and set to zero all  the mode functions with $n>N$.

The conservation laws are not affected by the truncation, except the energy conservation law  in the special case of the truncation at order $N=0$. These conservation laws are easily deduced by integrating the kinetic equation with suitable weights. 
The conservation law of the number density reads
\begin{align}
\frac{\partial {\cal N}}{\partial \tau} = -\frac{\cal N}{\tau} .
\label{num_conservation}
\end{align} 
It involves only the $f_0$ mode function, and its form is not modified in any truncation. In contrast, the conservation law of the local energy density
\begin{align}
\frac{\partial \mathcal{E}}{\partial \tau} = -\frac{\mathcal{E} +\mathcal{P}_L}{\tau} 
\end{align} 
involves also the $f_1$ mode through the longitudinal pressure $\mathcal{P}_L$. In the truncation orders $N\geq 1$, its form is not affected. However, in the truncation order $N=0$, the first order moment (\textit{i.e.} the pressure anisotropy) is forced to vanish, implying that the energy density and the pressure are always related as $\mathcal{P}_L=\mathcal{P}_T=\mathcal{E}/3$ (for massless particles). In this case, the conservation law takes the form
\begin{equation}
\frac{\partial \mathcal{E}}{\partial \tau} = -\frac{4}{3} \frac{\mathcal{E}}{\tau} 
\hspace{10pt} (\text{for } N=0) .
\end{equation}
This is nothing but the ideal hydrodynamic equation, whose solution is $\mathcal{E} (\tau) =\mathcal{E} (\tau_0) \left( \tau_0/\tau \right)^{4/3}$. Of course, this is also the regime expected when all   higher mode functions $n=1,2,\cdots$ have been damped by the collisions, and where the evolution of the system is fully described  by the zeroth mode function. 

%%%%%%%%%%%%%%%%%%%%%%%%%%%%%%%%%%%%%%%%%%%%%%%%%%%%%%%%%%%%%%%%%%%%%%%%%%%%%%%%%%%%%%%
\section{Initial conditions and time scales} \label{sec:initial}
%%%%%%%%%%%%%%%%%%%%%%%%%%%%%%%%%%%%%%
\begin{figure}[tb]
 \begin{tabular}{cc}
 \begin{minipage}{0.5\hsize}
  \begin{center}
   \includegraphics[clip,width=8cm]{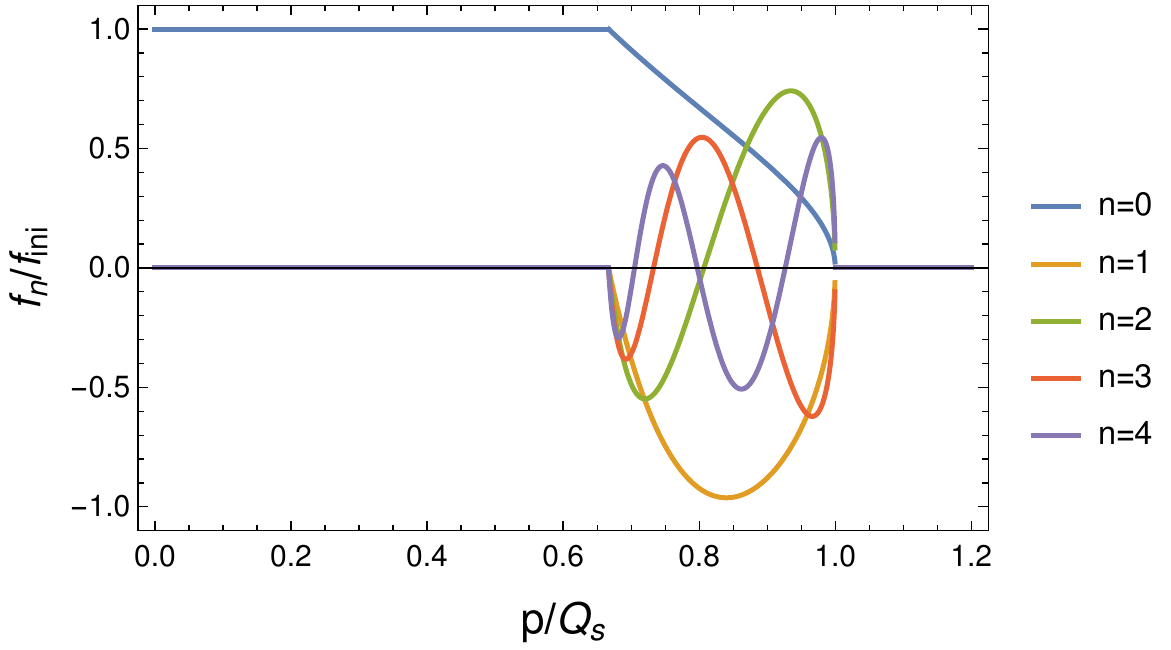}
  \end{center}
 \end{minipage} &
 \begin{minipage}{0.5\hsize}
  \begin{center}
   \includegraphics[clip,width=8cm]{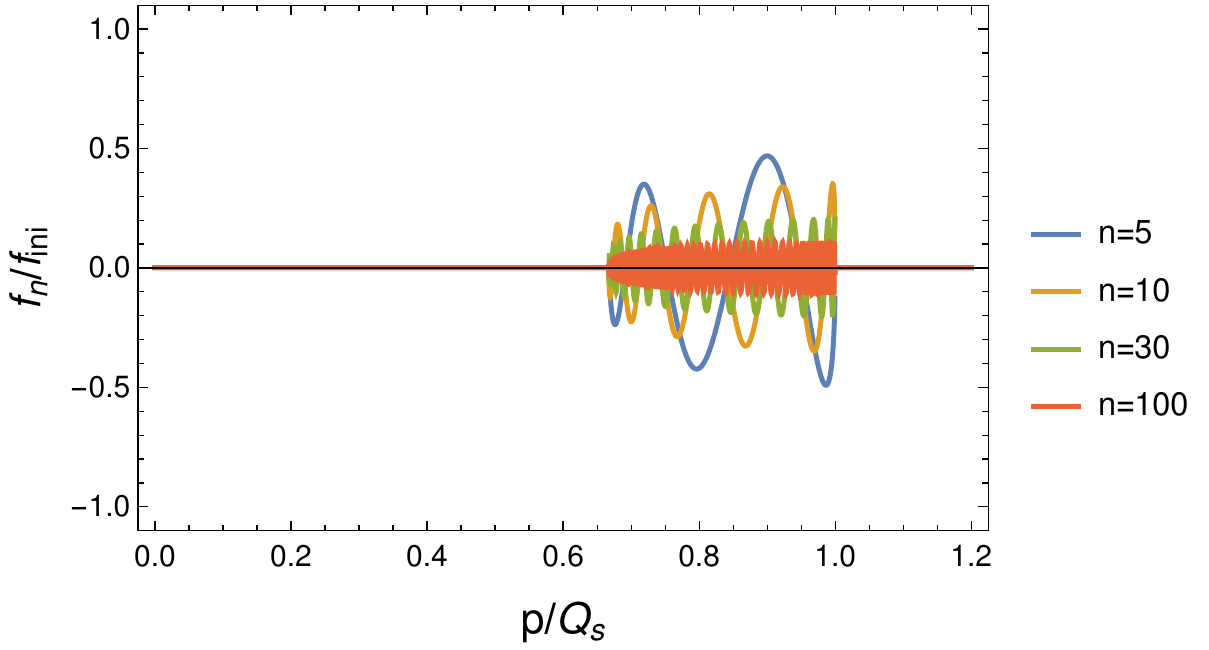}
  \end{center}
 \end{minipage} 
 \end{tabular}
 \caption{Initial mode functions \eqref{ini_mode} for several $n$. The anisotropy parameter is $\xi_0=1.5$.}
 \label{fig:ini_fn}
\end{figure}

Before we present some numerical results, we discuss shortly the initial conditions used in the calculations. We also provide an estimate of the time scale that   controls the collision rate. 

\subsection{Initial conditions} \label{subsec:initial}
In a collision of high-energy nuclei, gluons whose typical momentum scale is given by the saturation momentum $Q_s$ are emitted in the central rapidity region \cite{Mueller:1999fp}. As in previous related studies (see e.g. \cite{Blaizot:2017lht}), 
we model the distribution of these gluons by the following initial distribution,
\begin{align}
f(\tau_0 ,\bp ) &= f_{\rm ini} \, \Theta \left( Q_s -\sqrt{(\xi_0 p_z )^2 +p_\perp^2}\right) \, , \label{f_ini}
\end{align}
where $\Theta (z)$ is the Heaviside step function, $f_{\rm ini}$ is a parameter that characterizes the initial occupation number, and $\xi_0>1$ measures the initial anisotropy. This particular form allows for a simple determination of the initial mode functions. We emphasize that none of our main conclusions depends crucially on the shape of this initial distribution. 
The time scale for the formation  of the initial  gluons is $\tau_0 \sim Q_s^{-1}$, and throughout we shall take this time as the starting time for the kinetic  evolution. Note that $Q_s$ is the only momentum scale, and we shall consistently express all momenta in units of $Q_s$. 

The initial conditions for the mode functions $f_n (p)$ that correspond to the initial distribution \eqref{f_ini} are given by
\begin{align}
f_n (\tau_0 ,p) 
&= \frac{4n+1}{2} f_{\rm ini} \int_{-1}^1 dx \, \Theta \left( Q_s -p \sqrt{(\xi_0^2 -1)x^2 +1}\right) P_{2n} (x) \notag \\
&= f_{\rm ini} \, \delta_{n,0} \Theta ( Q_s /\xi_0 -p) \notag \\
&
+f_{\rm ini} \left[ P_{2n+1} \left(\sqrt{\frac{Q_s^2-p^2}{(\xi_0^2-1)p^2}} \right) -(1-\delta_{n,0}) P_{2n-1} \left(\sqrt{\frac{Q_s^2-p^2}{(\xi_0^2-1)p^2}} \right) \right] \Theta (Q_s-p) \Theta (p-Q_s/\xi_0).
\label{ini_mode}
\end{align}
These are plotted in Fig.~\ref{fig:ini_fn} for several values of $n$ and an anisotropy parameter $\xi_0=1.5$. The mode functions for $n\geq 1$ are nonzero only in the region $Q_s/\xi_0 <p<Q_s$, where the distribution shows anisotropy. They exhibit oscillations whose wave number grows with increasing $n$. 

\begin{figure}[tb]
 \begin{tabular}{cc}
 \begin{minipage}{0.5\hsize}
  \begin{center}
   \includegraphics[clip,width=8cm]{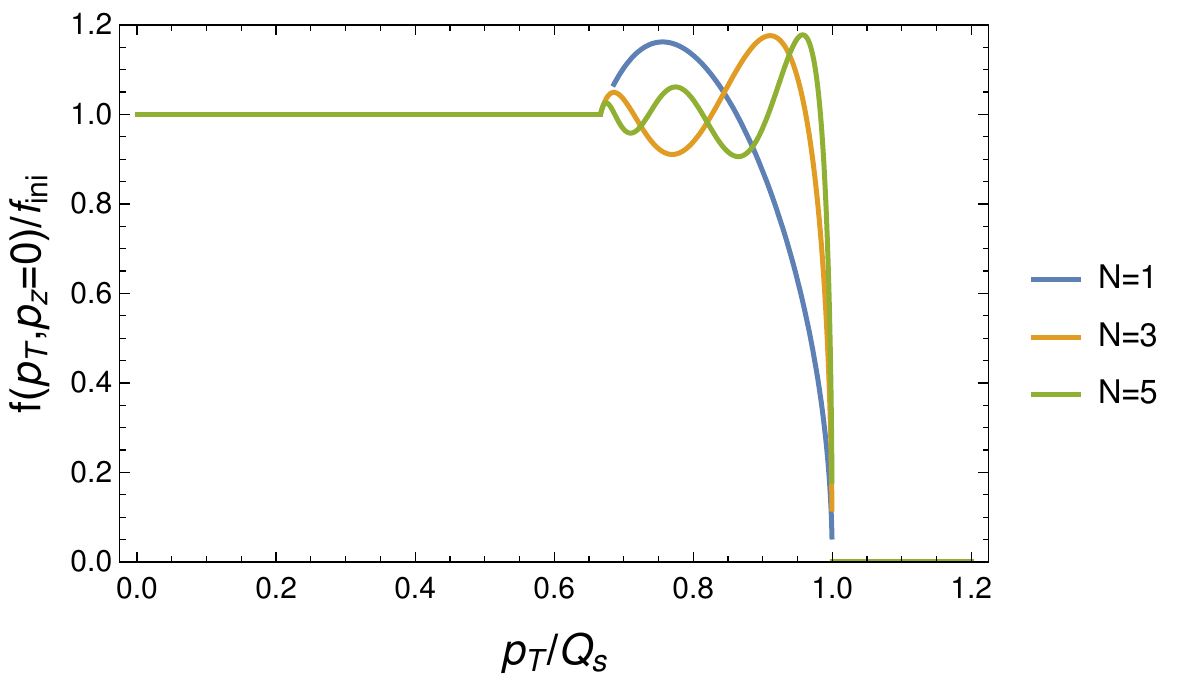}
  \end{center}
 \end{minipage} &
 \begin{minipage}{0.5\hsize}
  \begin{center}
   \includegraphics[clip,width=8cm]{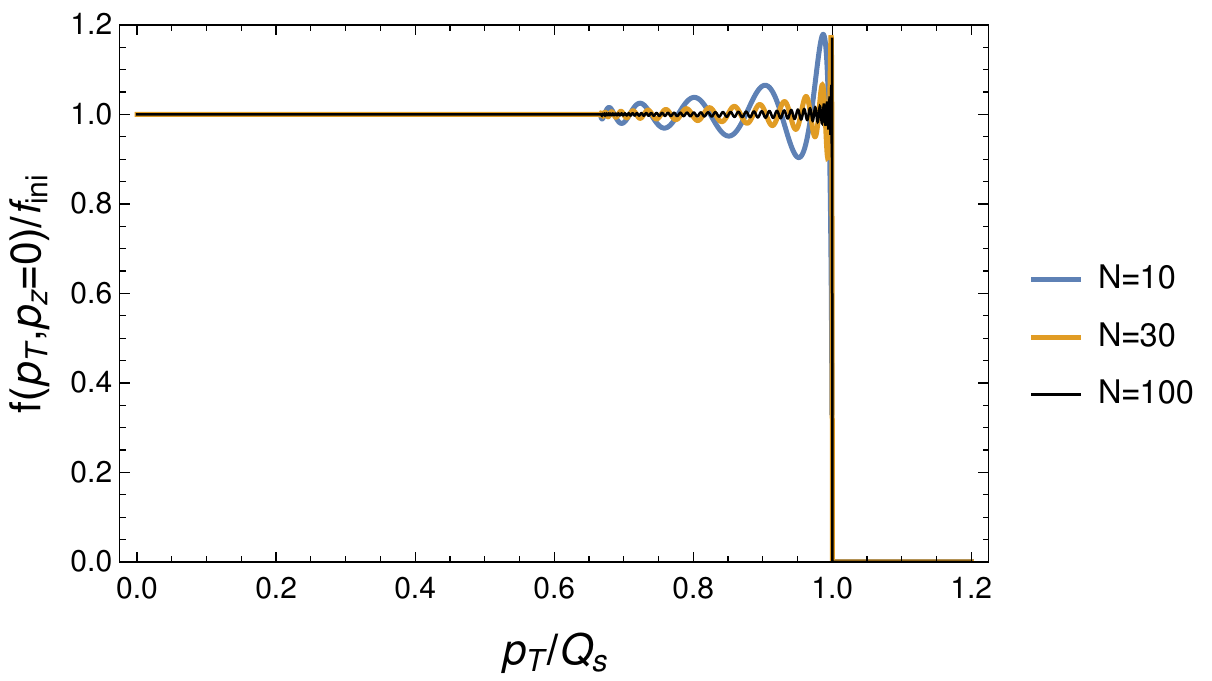}
  \end{center}
 \end{minipage} 
 \end{tabular}
 \caption{Reconstruction of the initial distribution with the truncated modes up to $N$. The distributions at $p_z=0$ are plotted as a function of $p_\perp$. }
 \label{fig:ini_reconst}
\end{figure}

In Fig.~\ref{fig:ini_reconst}, we show  how the original distribution is reconstructed by summing up the mode functions for several truncation orders. More precisely, we plot $\sum_{n=0}^N f_n (t,p) P_{2n} (\ct) $ at $p_z=0$ ($\cos \theta=0$) for several $N$. For large $N$, this sum indeed approaches the original step function. Not surprisingly,  the convergence is  slower at the corner of the distribution, near the sharp discontinuity at $p=Q_s$. However, we shall see  that such small structures in the distribution functions are quickly smeared out by collisions.

%%%%%%%%%%%%%%%%%%%%%%%%%%%%%%%%%%%%%%
\subsection{Time scales}

To analyze the results of the numerical calculations to be presented in the next  section, it is useful to measure the time in units of a quantity that can be associated to a basic  relaxation time $\tau_R$. This  time can be roughly estimated by comparing the collision term \eqref{col_sa0} with a relaxation time approximation, $(f-f_{\rm eq})/\tau_R$.  From this comparison, we find
\begin{equation}\label{estimatetauR0}
\frac{1}{\tau_R} \sim 4\pi N_c^2 \alpha_s^2 \, l_{\rm Cb}\, \frac{I_a}{p^2} \, .
\end{equation}
The combination $4\pi N_c^2 \alpha_s^2 \, l_{\rm Cb}$ is that already encountered in the  definition of $\tilde\tau$ in Eq.~(\ref{tildetau}). 
Thus defined, the relaxation time is a momentum and also time-dependent quantity. We fix  the momentum dependence by  evaluating $\tau_R$  at the typical momentum scale of the gluons  at the early time, \textit{i.e.} $p=Q_s$. 
The time dependence comes from the integral $I_a$ defined in Eq.~\eqref{Ia}. If the typical occupation number is not much larger than unity, one can neglect the Bose-enhancement factor.\footnote{We shall not consider in this paper the case of large occupation, $f_{\rm ini}\sim 1/\alpha_s$, which has been thoroughly analyzed in Ref.~\cite{Tanji:2017suk}.} (We typically use for initial occupation number $f_{\rm ini}=0.1$.) In that case, the integral $I_a$  is approximately equal to  the number density ${\cal N}$ (modulo the factor $\nu_\text{g}$) and, for the expanding case where ${\cal N}(\tau)\propto 1/\tau$, it can be evaluated as
\begin{equation}
I_a (\tau) \simeq \frac{\cal N}{\nu_\text{g}} = \frac{Q_s^3 f_{\rm ini}}{6\pi^2 \xi_0} \frac{1}{Q_s \tau}. 
\end{equation}
The relaxation time for the expanding case is therefore estimated as
\begin{equation}\label{tauR}
\tau_R (\tau) \sim \frac{3\pi \xi_0}{2N_c^2 \alpha_s^2 \,l_{\rm Cb} \, f_{\rm ini}} \tau \, .
\end{equation}
Since the system gets diluted proportionally to  $\tau$ by the longitudinal expansion, the expansion rate decreases as $1/\tau$. With the present estimate of $\tau_R(\tau)$, the ratio between the collision rate and the expansion rate is therefore independent of $\tau$. This is an oversimplification however. In the following we shall present results in terms of a time variable measured in units of the relaxation time $\tau_R$ defined by the expression (\ref{tauR}) at the initial time $\tau_0$, that is 
\begin{equation}\label{tauR0}
\tau_R \equiv  \frac{3\pi \xi_0}{2N_c^2 \alpha_s^2 \,l_{\rm Cb} \, f_{\rm ini}} \tau_0 \, .
\end{equation}
By doing so, we do not need to specify the values of the various parameters entering the definition (\ref{tauR0}), in particular the coupling constant $\alpha_s$. Note however that for typical values $\alpha_s\simeq 0.3$, $\xi_0\simeq 1$, $l_{\rm Cb} \simeq 1$, one finds $\tau_0/\tau_R\simeq 0.2 f_{\rm ini}$,  as a crude order of magnitude. 

In the next section, we shall also consider the non expanding case. In this case, the density is constant, and so is $I_a$. The relaxation time is then independent of time and is defined as in Eq.~(\ref{tauR0}), with $\tau_0$ replaced by $1/Q_s$, 
\beq\label{tauRstatic}
\tau_R=\frac{3\pi \xi_0}{2N_c^2 \alpha_s^2 \,l_{\rm Cb} \, f_{\rm ini}} \frac{1}{Q_s}.
\eeq

%%%%%%%%%%%%%%%%%%%%%%%%%%%%%%%%%%%%%%%%%%%%%%%%%%%%%%%%%%%%%%%%%%%%%%%%%%%%%%%%%%%%%%%
\section{Numerical results} \label{sec:results}
In this section, we show numerical results obtained by solving the kinetic equations \eqref{Boltz_n} for the mode functions  with the initial conditions \eqref{ini_mode}. To show the efficiency of the angular mode expansion method, we compare these results with those obtained by solving   the full Boltzmann equation, Eqs.~\eqref{Boltz0} and \eqref{col_sa0}, using the method presented in Ref.~\cite{Tanji:2017suk}. Since the kinetic equations for the individual mode functions take almost the same form as a Boltzmann equation for an isotropic system, they can be solved by the same method as the original Boltzmann equation but with much lower numerical cost, since they are one-dimensional instead of two-dimensional.

Below we employ a fixed set of parameters for the initial conditions $f_{\rm ini}=0.1$ and $\xi_0 =1.5$. In numerical calculations, we have nonuniformly discretized the momentum variable $p\in [p_\text{min} ,p_\text{max}]$ into 200--400 grid points, where $p_\text{max}/Q_s=4$--8 and $p_\text{min}/Q_s=10^{-2}$. The time evolution is solved by an implicit method.

We shall first consider the non expanding case and analyze how the system isotropize and thermalize. Then we turn to the expanding case, and consider the same issues for various values of the collision rate. 

%%%%%%%%%%%%%%%%%%%%%%%%%%%%%%%%%%%%%%
\subsection{Non-expanding plasma} \label{subsec:fixedbox}

For a non-expanding plasma, we need to solve the following  set of kinetic equations
\begin{equation}
\frac{\partial}{\partial t} f_n (p) = -C_n [f] \, ,
\end{equation}
where the drift terms ($\propto 1/\tau$) in Eq.~(\ref{Boltz_n})  have been left aside. We shall do so by truncating this infinite set of equations to a finite number, $n\le N$. We shall see that values $N\le 4$ are actually sufficient in order to get accurate results (in most cases, even $N=1$ is enough). The results obtained with this method will be compared with the solution of the corresponding kinetic equation solved by the method of Ref.~\cite{Tanji:2017suk}.
In the non-expanding system, the value of the initial time does not affect the results, so we chose $t_0=0$. We shall express all results as a function of $t/\tau_R$, where $\tau_R$ is the relaxation time defined in Eq.~(\ref{tauRstatic}). 

Both  particle number and  energy are conserved to an accuracy better than  $10^{-3}$  within the time range shown in the following figures.

\begin{figure}[H]
 \begin{center}
  \includegraphics[clip,width=7.5cm]{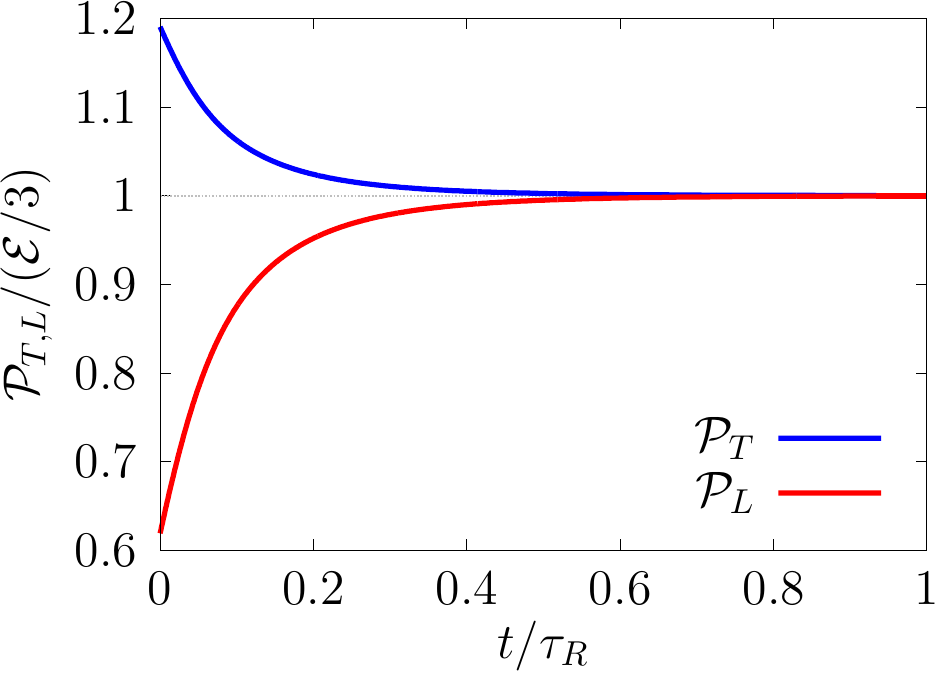}\qquad 
  \includegraphics[clip,width=7.5cm]{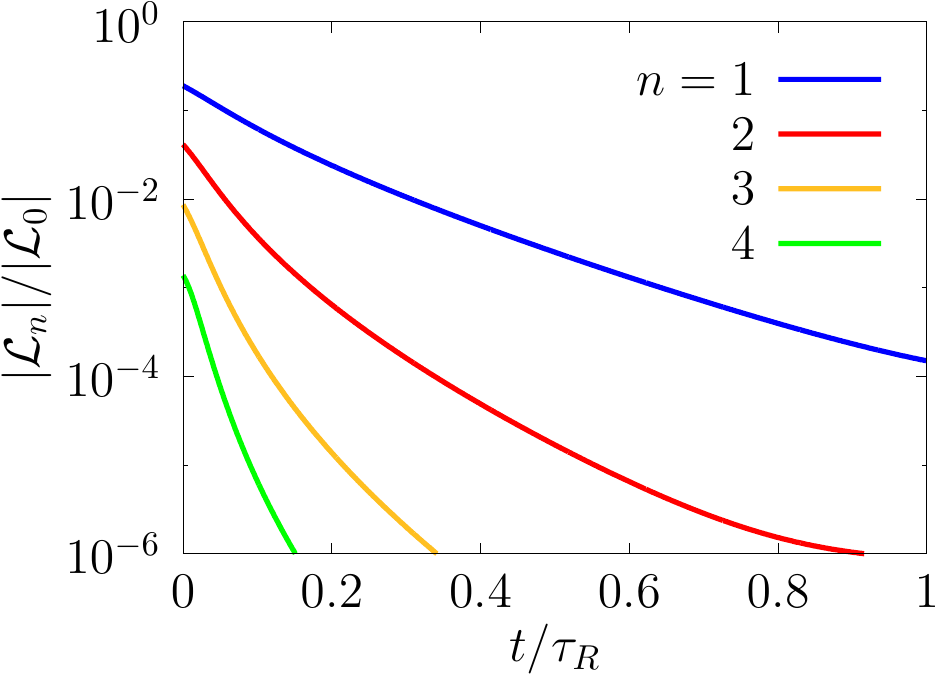}
  \vspace{-20pt}
 \end{center}
 \caption{Left: time evolution of the transverse and longitudinal pressures, normalized to $\mathcal{E}/3$. Right: time evolution of  the moment $\mathcal{L}_n$ for $n=1,2,3,4$ (from top to bottom) divided by the zeroth moment $\mathcal{L}_0$. }
 \label{fig:pressure_FB}
\end{figure}
For a first orientation, we show in Fig.~\ref{fig:pressure_FB}, left panel, the longitudinal and the transverse pressures divided by the energy density as a function of time. 
One sees that  the pressures are isotropized on a  time scale $\lesssim \tau_R$. 
The right panel of Fig.~\ref{fig:pressure_FB} shows the time evolution of the moments $\mathcal{L}_n$ defined in Eq.~\eqref{def_momL}. As was observed already in Ref.~\cite{Blaizot:2017lht}, there is a hierarchy of the moments $|\mathcal{L}_0| > |\mathcal{L}_1| >|\mathcal{L}_2| >\cdots$, and all these moments  decay quickly in time,  the higher the rank $n$ the faster the decay. This is another and more detailed way to visualize the  isotropization of the system as a function of time. The results in Fig.~\ref{fig:pressure_FB} were obtained by solving the full Boltzmann equation. However as we shall see now, equivalent results are obtained with  low order truncations.

\begin{figure}[H]
 \begin{center}
  \includegraphics[clip,width=7.5cm]{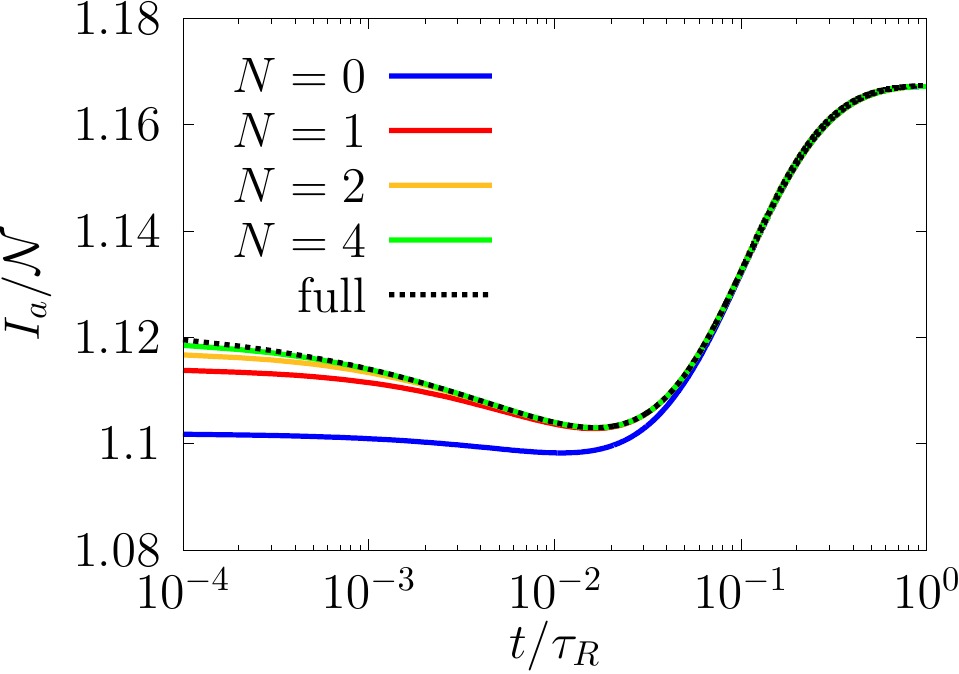}\qquad  
  \includegraphics[clip,width=7.5cm]{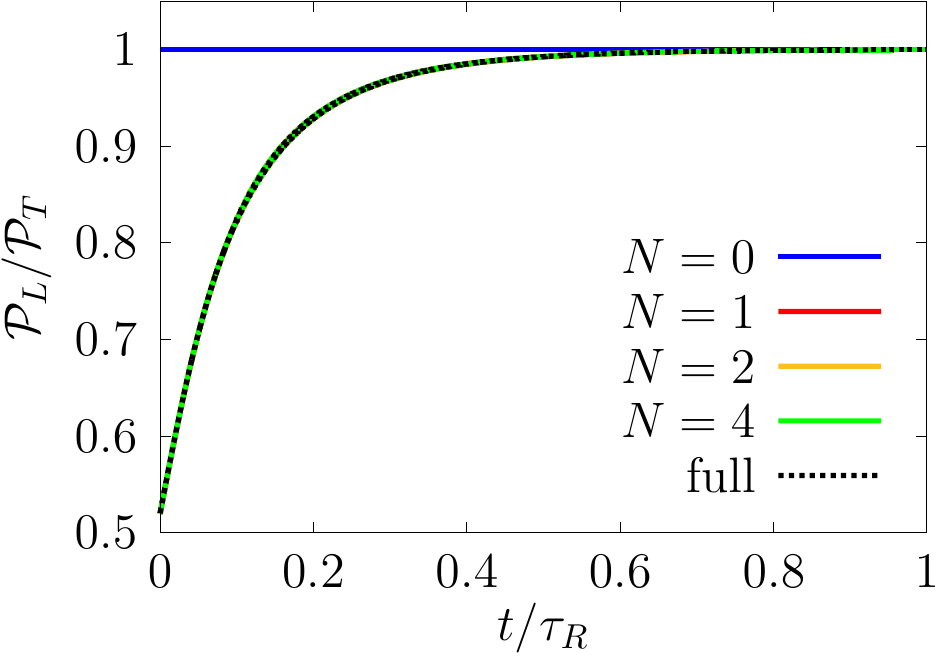}
  \vspace{-20pt}
 \end{center}
 \caption{Left: time evolution of $I_a$ (divided by the number density ${\cal N}$). Several truncations $N=0,1,2,4$ are compared to the full result indicated by the dotted line. Right: time evolution of the pressure ratio $\mathcal{P}_L/\mathcal{P}_T$, for $N=0,1,2,4$ and that of the full result. All the lines for $1\le N$ are indistinguishable for $t/\tau_R\gtrsim 10^{-2}$, showing the excellent convergence of the expansion.}
 \label{fig:Ia_FB}
\end{figure}

In the non-expanding system, the different angular modes couple to each other only through the terms nonlinear  in $f$ that are present in the collision kernel of Eq.~(\ref{coll_n}). One of such nonlinearity is hidden in the integral $I_a$ given in Eq.~\eqref{Ia_mode}. The time evolution of $I_a$ (divided by the number density $\cal N$) is plotted for several truncation orders $N\leq 4$ in Fig.~\ref{fig:Ia_FB}, left panel. For comparison, the result obtained from the full Boltzmann equation  is also plotted as a dotted line. The nonlinearity is measured here by the deviation of $I_a/{\cal N}$ from unity (when $f\ll 1$, $I_a\simeq {\cal N}$).  Small deviations from the full result are seen only at early times and rapidly disappear in  higher order truncations. At later time, even the 0th order truncation agrees with the full result. The right panel of Fig.~\ref{fig:Ia_FB} displays the time evolution of the pressure ratio $\mathcal{P}_L/\mathcal{P}_T$. Note that in the lowest truncation order $N=0$, the anisotropy is entirely neglected and the pressure ratio is always one. Truncations with $N\geq 1$ fully account for the anisotropy of pressures. As seen in the figure, all the results for $N=1,2,4$ nicely agree with the full result. This indicates that the coupling to higher modes $n=2,3,\cdots$ does not play much of a role in the time evolution of the pressures: it is enough to keep the lowest angular modes $f_0$ and $f_1$ to get an accurate description. 

\begin{figure}[H]
 \begin{tabular}{cc}
 \begin{minipage}{0.5\hsize}
  \begin{center}
   \includegraphics[clip,width=7.5cm]{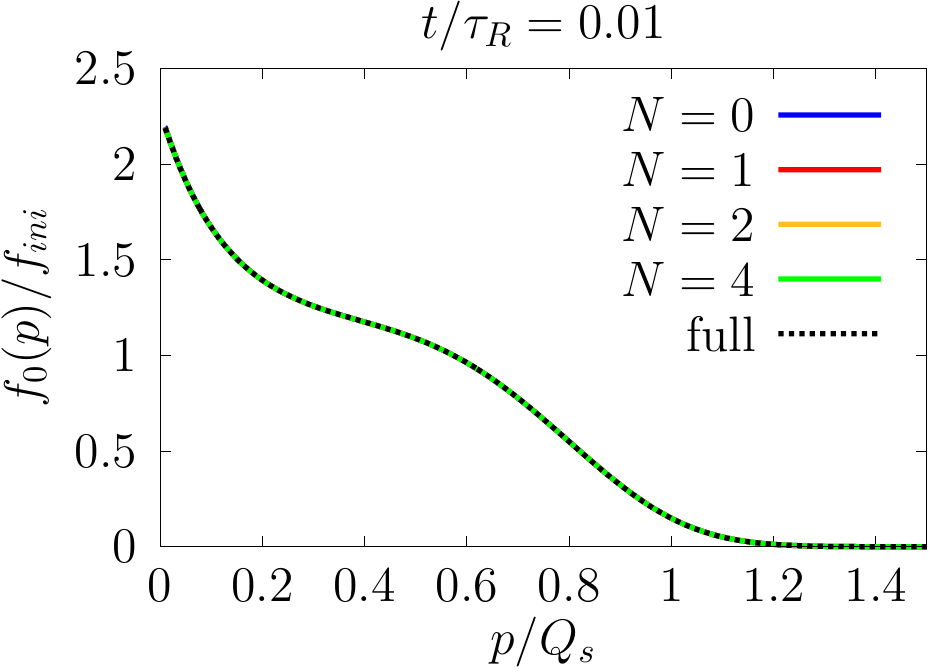}
  \end{center}
 \end{minipage} &
 \begin{minipage}{0.5\hsize}
  \begin{center}
   \includegraphics[clip,width=7.5cm]{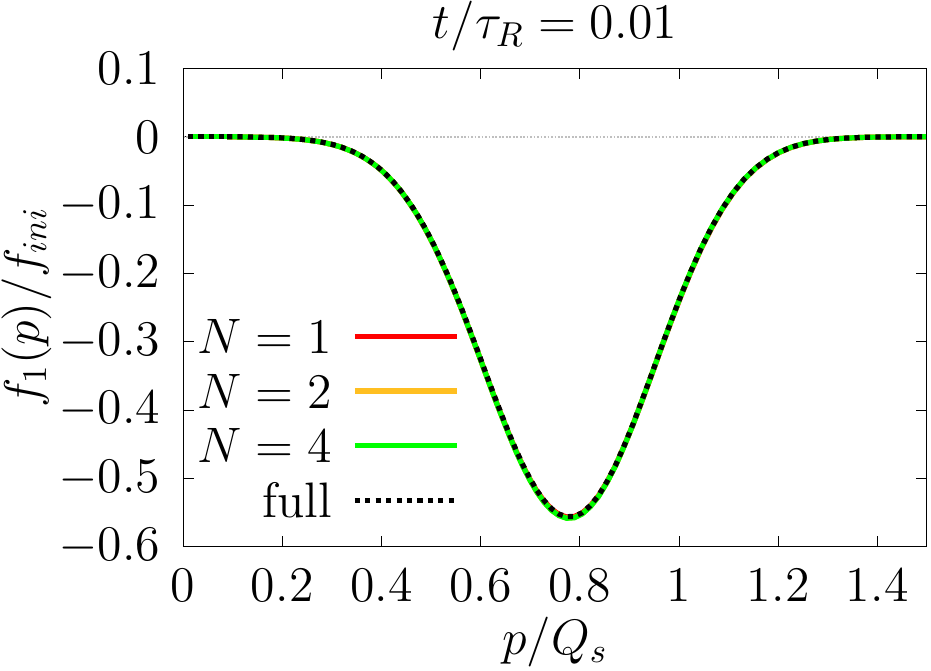}
  \end{center}
 \end{minipage} 
 \end{tabular}
 \caption{Mode functions $f_0(p)$ (left) and  $f_1(p)$ (right) at time $t/\tau_R =0.01$ obtained with  truncation orders $N\le 4$ and compared to the full result. All the lines completely overlap with each other.}
 \label{fig:Ndep_f_FB}
\end{figure}

To have a closer look at the efficiency of the truncation, we plot in Fig.~\ref{fig:Ndep_f_FB} the first two mode functions $f_0(p)$ and $f_1(p)$ as a function of momentum at some early time $t/\tau_R=0.01$,  for several truncation orders. The results corresponding to different truncations completely overlap with each other, and are in perfect agreement with the full result represented by the dotted line. Thus, already at  early time ($t\ll\tau_R$) the truncation of the angular modes works extremely well. Note that this does not necessarily imply that all the other angular modes are vanishingly small. Simply they do not significantly affect the dynamics of the lowest modes, in which we are primarily interested. \\

\begin{figure}[H]
 \begin{tabular}{cc}
 \begin{minipage}{0.5\hsize}
  \begin{center}
   \includegraphics[clip,width=7.5cm]{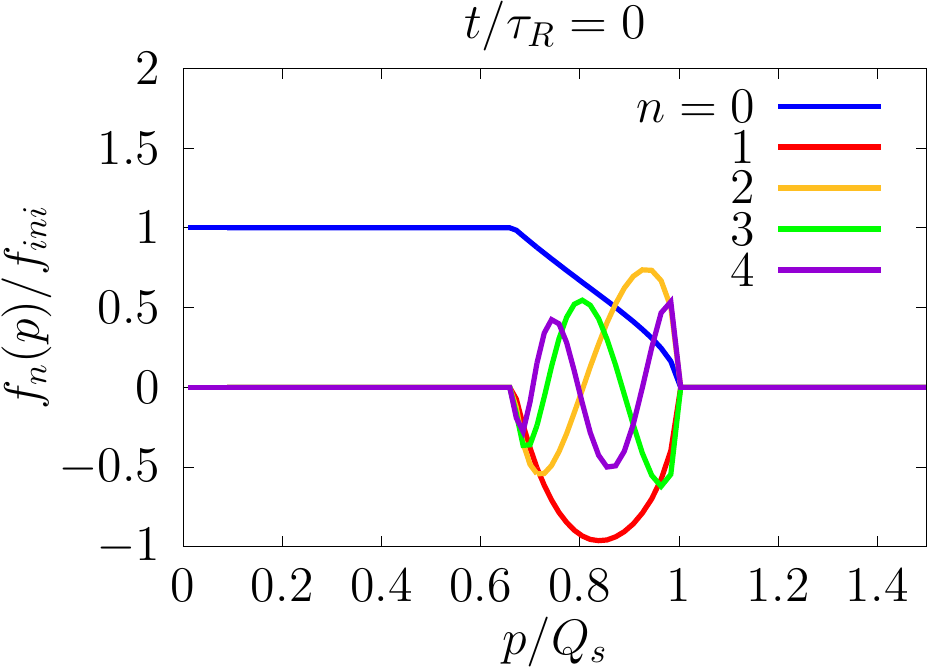}
  \end{center}
 \end{minipage} &
 \begin{minipage}{0.5\hsize}
  \begin{center}
   \includegraphics[clip,width=7.5cm]{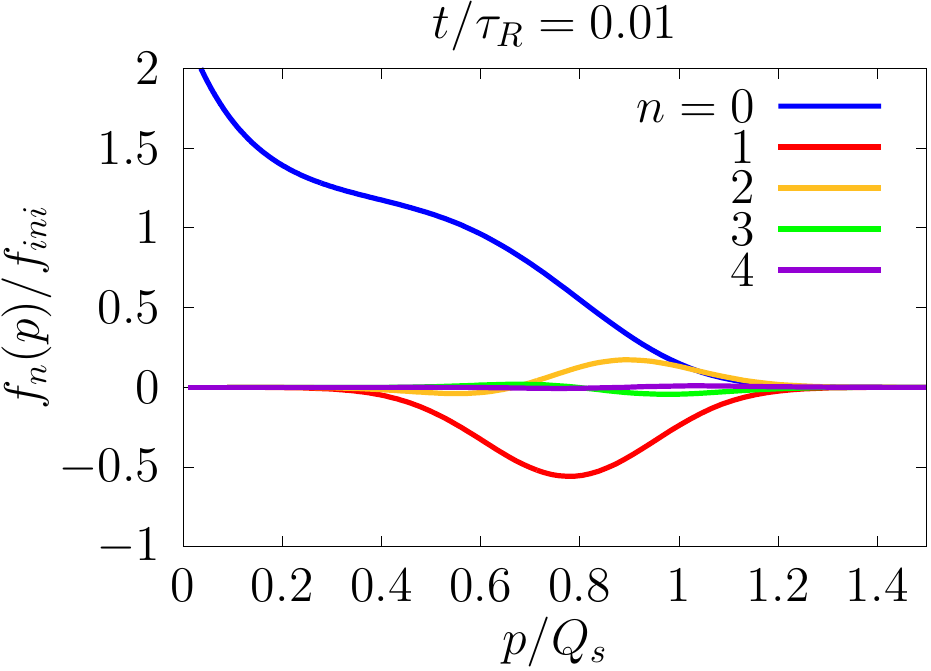}
  \end{center}
 \end{minipage} \\
 \ & \\[-5pt]
 \begin{minipage}{0.5\hsize}
  \begin{center}
   \includegraphics[clip,width=7.2cm]{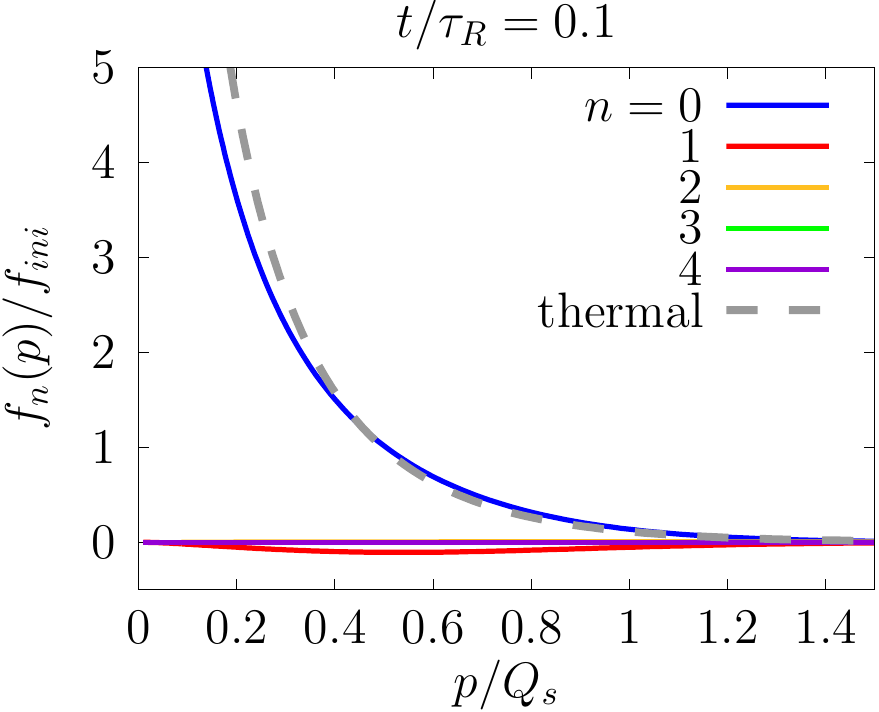}
  \end{center}
 \end{minipage} &
 \begin{minipage}{0.5\hsize}
  \begin{center}
   \includegraphics[clip,width=7.2cm]{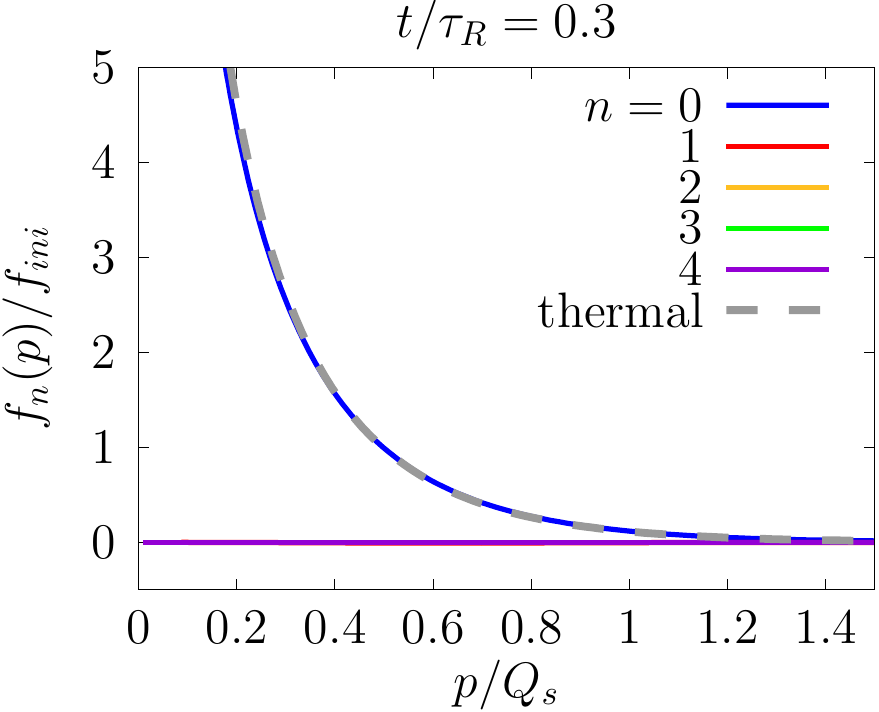}
  \end{center}
 \end{minipage} 
 \end{tabular}
 \caption{Mode functions $f_n(p)$ for $n=0,1,2,3,4$ at times $t/\tau_R=0$, 0.01, 0.1, 0.3 computed in the truncation order $N=4$. In the lower two panels, the thermal Bose-Einstein distribution with $T=0.23Q_s$ and $\mu=-0.075Q_s$ is also plotted as a dashed line. }
 \label{fig:fn_FB}
\end{figure}

We can push this analysis a bit further and  make a more detailed comparison of the evolution of the mode functions $f_n(p)$ for different $n$. In Fig.~\ref{fig:fn_FB}, the mode functions for $n=0,1,2,3,4$ at times $t/\tau_R=0$, 0.01, 0.1, 0.3 are plotted as a function of momentum. These results are obtained with the truncation $N=4$. As we have already seen in Sec.~\ref{subsec:initial}, the angular modes $n\geq 1$ exhibit initially rapid oscillations. However,  these oscillations are quickly washed out, as seen in the right upper panel ($t/\tau_R=0.01$). As time goes on, all the higher mode functions decay in the whole momentum region and only the mode  $f_0$ survives at a later time.
In the lower two panels, the thermal distribution $f_{\rm eq} (p)=1/(e^{(p-\mu)/T}-1)$ is also plotted as a gray dashed line. In the non-expanding system, with conserved particle number, one can calculate the temperature $T$ and chemical potential $\mu$ of the final state from the initial condition. For our initial condition, these are $T=0.23\,Q_s$ and $\mu=-0.075\,Q_s$. From  $t/\tau_R \simeq 0.3$, when all the higher mode functions have essentially decayed, the zero mode function nicely agrees with the thermal distribution. 

\begin{figure}[H]
 \begin{tabular}{cc}
 \begin{minipage}{0.5\hsize}
  \begin{center}
   \includegraphics[clip,width=7.5cm]{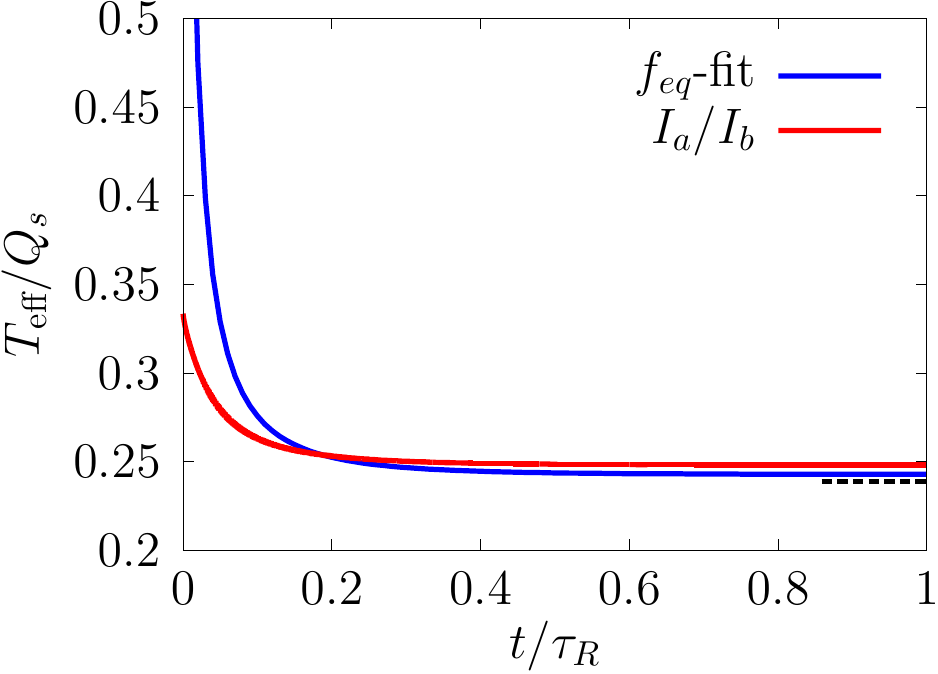}
  \end{center}
 \end{minipage} &
 \begin{minipage}{0.5\hsize}
  \begin{center}
   \includegraphics[clip,width=7.5cm]{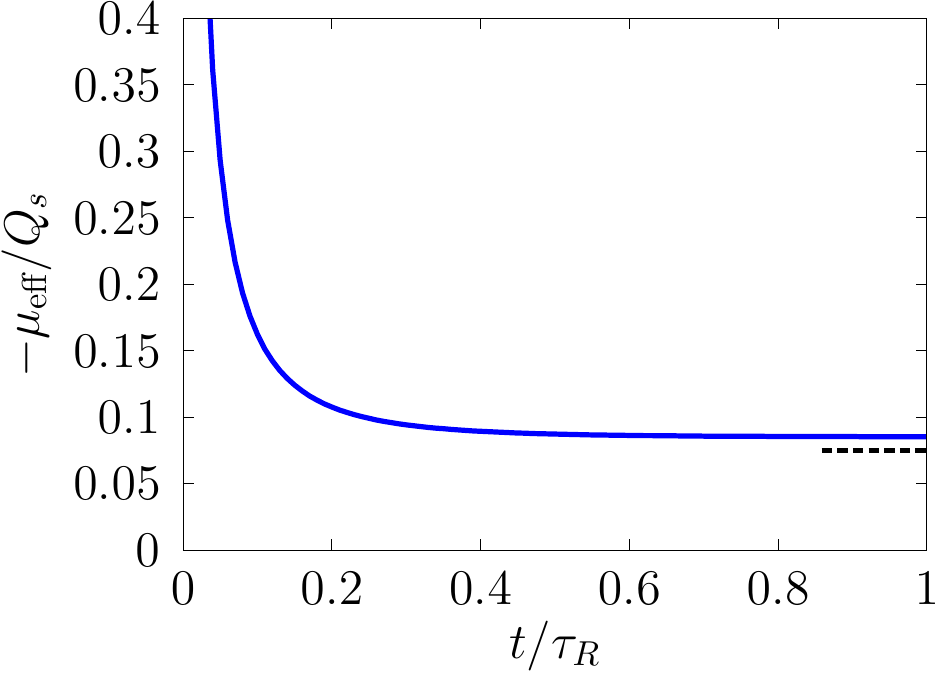}
  \end{center}
 \end{minipage} 
 \end{tabular}
 \caption{Time evolution of the effective temperature and chemical potential in the non-expanding system. For the effective temperature, two different methods are compared (see text). The final equilibrated values calculated by the conservation laws are indicated by black dashed lines.}
 \label{fig:Tmueff_FB}
\end{figure}

\begin{figure}[H]
 \begin{center}
  \includegraphics[clip,width=8cm]{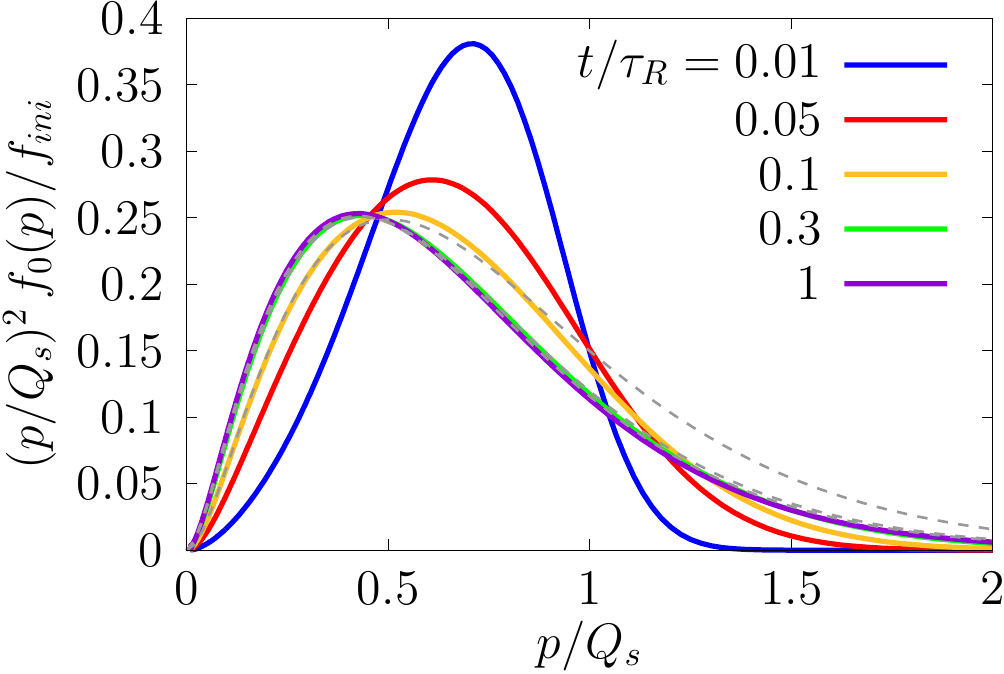}
  \vspace{-20pt}
 \end{center}
 \caption{Time evolution of $f_0 (p)$ multiplied by $p^2$, computed in the truncation order $N=4$. The dashed lines represent thermal distributions, at $t/\tau_R=0.1$, 0.3, 1, with the effective temperature and chemical potential  determined as explained in the text (the curves with $t/\tau_R=0.3$ and 1 are indistinguishable on the figure). Note that for $t=0.1 \tau_R$, the higher momentum modes with $p\gtrsim Q_s$ are underpopulated as compared to the thermal distribution. }
 \label{fig:tdep_f0_FB}
\end{figure}

The thermalization of the system involves not only the decay of the modes $f_n(p)$ for $n\ge 1$, a process that we may refer to as isotropization, but also the rearrangement of the various isotropic momentum modes, that is the evolution of $f_0(p)$ towards a thermal distribution, as we have just mentioned\footnote{Note that we do not find particular distinct time scales that could be attributed to each of these processes which rather progress in parallel.}. To study more precisely this phenomenon, we have extracted an effective temperature $T_\text{eff}$ and chemical potential $\mu_\text{eff}$ by fitting the zero mode function $f_0 (p)$ to an equilibrium distribution $f_{\rm eq} (p)=1/(e^{(p-\mu_\text{eff})/T_\text{eff}}-1)$ in the momentum range $0<p<Q_s$ (see e.g. \cite{Blaizot:2013lga} where a similar strategy has been used). The resulting values are plotted in Fig.~\ref{fig:Tmueff_FB} as a function of time. 
Except at early times $t/\tau_R \ltsim 0.1$, the fit is accurate, thus we do not indicate fitting errors in the figure. 
Both the effective temperature and the chemical potential converge toward their  expected values that can be deduced  from the conservation laws, and which are indicated by black dashed lines. 
For the effective temperature, we have employed also another definition $T_\text{eff}=I_a/I_b$, which coincides  with the actual temperature if the distribution function is an equilibrium distribution. Both definitions of the effective temperature yield nearly the same values at late time. As can be seen in Fig.~\ref{fig:tdep_f0_FB} the agreement of $f_0(p)$ with a thermal distribution is excellent  already at $t/\tau_R\simeq 0.3$. By that time, the pressures are nearly equilibrated, but not quite, the transverse pressure exceeding the  longitudinal pressure by a few percents (see Figs.~\ref{fig:pressure_FB} and \ref{fig:Ia_FB}).  Part of the reason is that, by this time the high momentum modes are not fully thermalized. This is obviously the case at $t/\tau_R=0.1$, as Fig.~\ref{fig:tdep_f0_FB} shows (the low momentum modes quickly thermalize).

%%%%%%%%%%%%%%%%%%%%%%%%%%%%%%%%%%%%%%
\subsection{Expanding plasma}

We turn now to the expanding case. Our goal here is twofold. On the one hand, we wish to verify that the convergence of the angular mode expansion remains of a good quality. And on the other hand, we want to verify the conclusions of another study, based on the analysis of a set of coupled equations for special moments, and a simplified collision kernel based on a relaxation time approximation  \cite{Blaizot:2017ucy,Blaizot:2019}. As shown in the study in question, the competition between free streaming and collisional effects can be understood from the competition between two fixed points that govern the evolution of the moments. The first fixed point is that of the free streaming motion, and it is independent of the collision kernel. The main property of this fixed point is that it is mostly determined by the two equations that control the evolution of the lowest moments ${\cal L}_0$ and ${\cal L}_1$, that is, in the present case, it involves only the two mode functions $f_0$ and $f_1$. The other moments, or mode functions, correct the position of the fixed point, but only in a moderate way, and most importantly they do not alter the qualitative behavior of the solution. The other fixed point corresponds to the hydrodynamic behavior to be expected at late time, unless the collision rate is too small as compared to the expansion rate. The hydrodynamics is  dominated by the first two mode functions, or equivalently the two moments ${\cal L}_0$ and ${\cal L}_1$ that contribute directly to the energy momentum tensor.

We shall start by discussing the case where the collision rate is comparable to the expansion rate, that is  $\tau_0 /\tau_R =1$. As we shall see, in this case, the collisions are nearly as efficient as in the non expanding case, they isotropize the system and drives it to local equilibrium. Then we discuss a few results obtained for smaller collision rates, where the evolution of the system becomes eventually dominated by the free streaming fixed point.  

\subsubsection{Large collision rate ($\tau_0 /\tau_R =1$)}

By large collision rate, we mean here that the collision rate is comparable to the expansion rate. The expansion plays an important role, but to a large extent the features that we have exhibited in the previous subsection for  the non-expanding case survive. 

\begin{figure}[H]
 \begin{center}
  \includegraphics[clip,width=7.5cm]{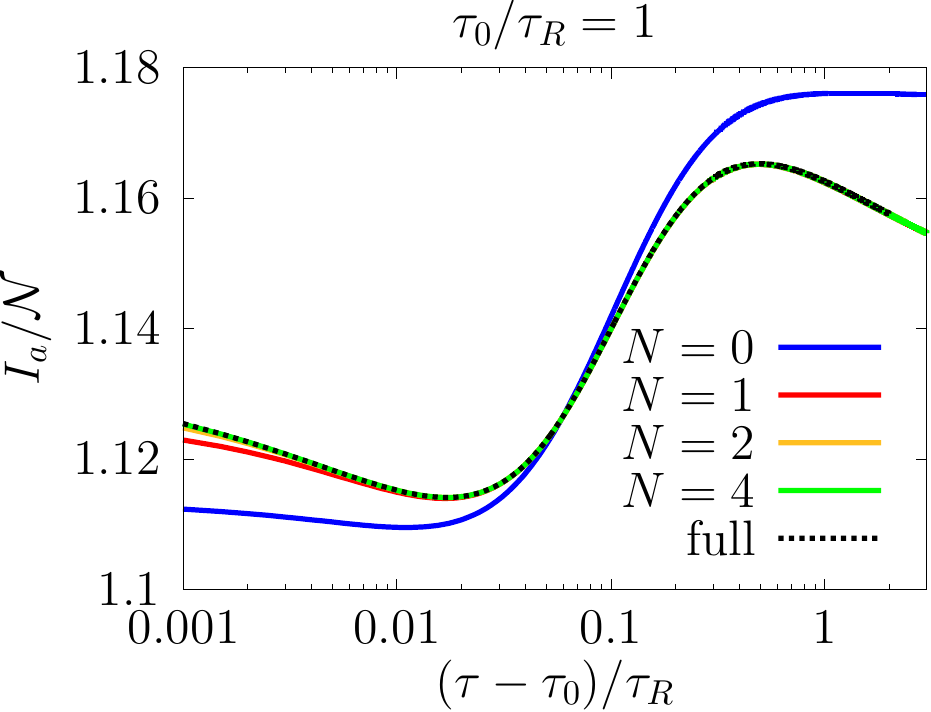}\qquad 
  \includegraphics[clip,width=7.5cm]{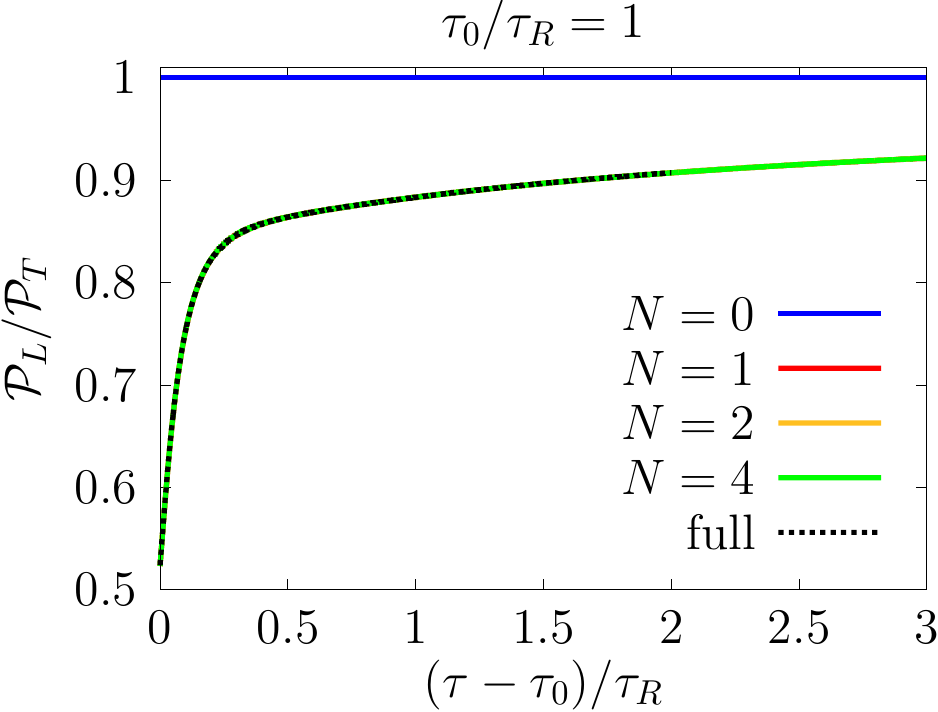}
  \vspace{-20pt}
 \end{center}
 \caption{Plot of the integral $I_a$ divided by the number density $\cal N$ (left), and of  the pressure ratio $\mathcal{P}_L/\mathcal{P}_T$ (right). Several truncations, $N=0,1,2,4$, are compared to the full result indicated by the dotted line.}
 \label{fig:Ia_exp2}
\end{figure}

The plots in Fig.~\ref{fig:Ia_exp2} reveal the quality of the truncations in the expanding case. It is to be compared to the corresponding plot for the non-expanding case in Fig.~\ref{fig:Ia_FB}:  the quality of the truncation is in both cases the same, and  the truncation $N=1$ becomes quickly accurate (note however the change of horizontal scale between the two panels, from logarithmic to linear).  

\begin{figure}[H]
 \begin{tabular}{cc}
 \begin{minipage}{0.5\hsize}
  \begin{center}
   \includegraphics[clip,width=7.5cm]{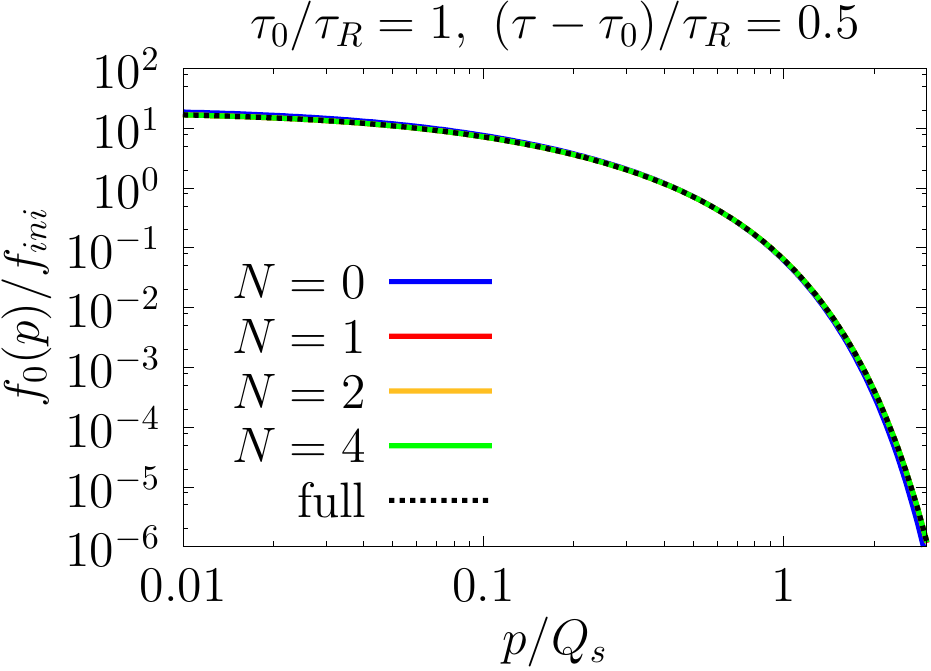}
  \end{center}
 \end{minipage} &
 \begin{minipage}{0.5\hsize}
  \begin{center}
   \includegraphics[clip,width=7.5cm]{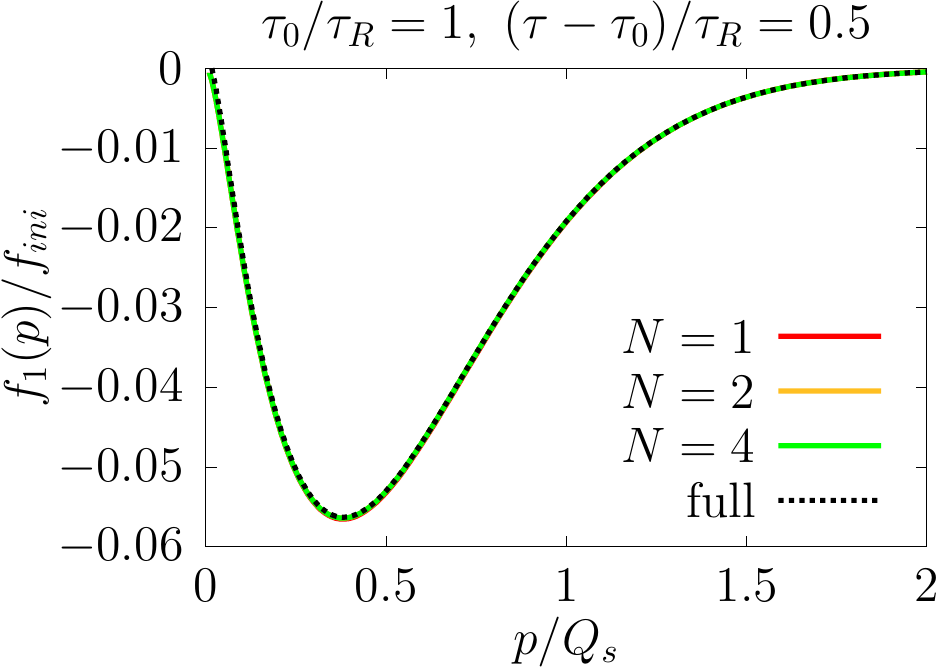}
  \end{center}
 \end{minipage} 
 \end{tabular}
 \caption{Mode functions $f_0(p)$ (left, log-log plot) and for $f_1(p)$ (right, linear plot) at time $(\tau-\tau_0)/\tau_R=0.5$ for several truncations. The dotted lines represent the exact solution of the Boltzmann equation.}
 \label{fig:Ndep_fn_Exp1}
\end{figure}

\begin{figure}[H]
 \begin{tabular}{cc}
 \begin{minipage}{0.5\hsize}
  \begin{center}
   \includegraphics[clip,width=7.5cm]{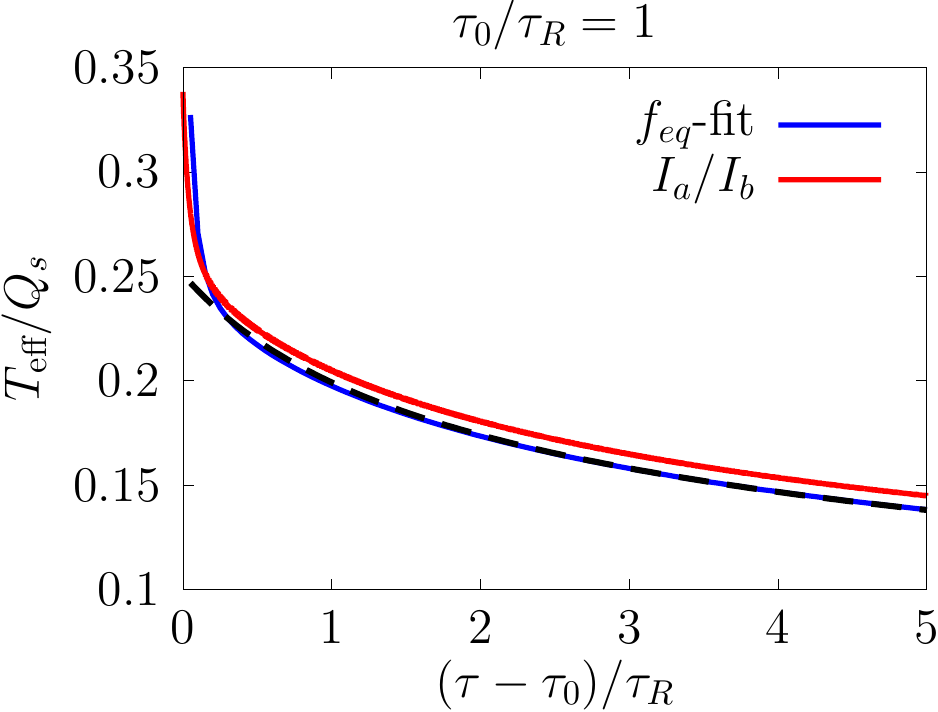}
  \end{center}
 \end{minipage} &
 \begin{minipage}{0.5\hsize}
  \begin{center}
   \includegraphics[clip,width=7.5cm]{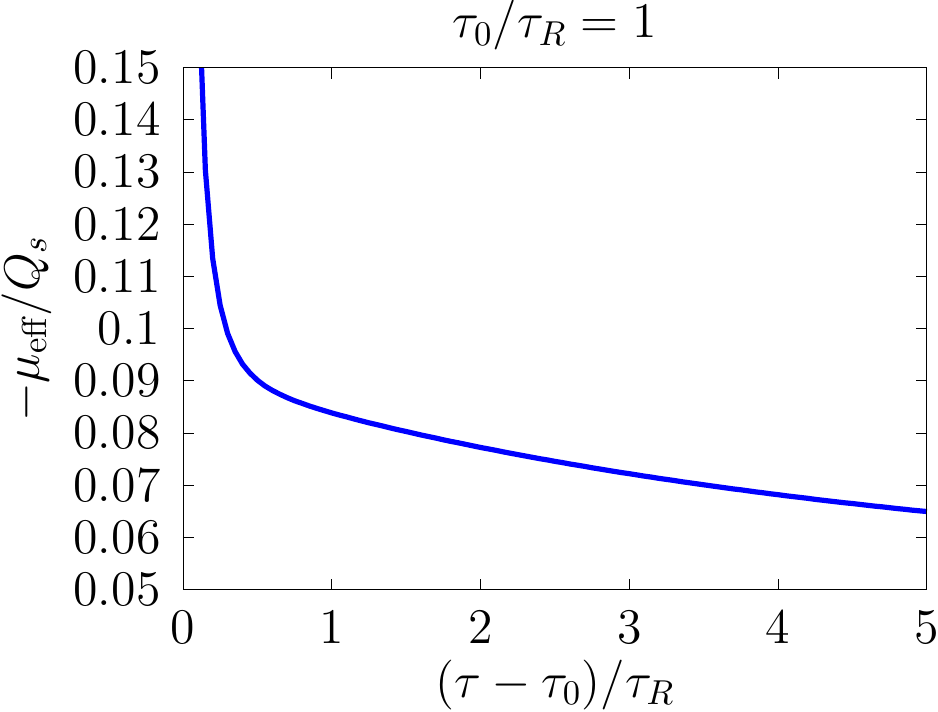}
  \end{center}
 \end{minipage} 
 \end{tabular}
 \caption{Time evolution of the effective temperature and chemical potential. The black dashed line in the left figure indicate the $\tau^{-1/3}$ behavior expected for ideal hydrodynamical evolution.}
 \label{fig:Tmueff_Exp1}
\end{figure}

The accuracy of the truncation is further confirmed by the analysis of the mode functions. As an example, we display in Fig.~\ref{fig:Ndep_fn_Exp1} the mode functions $f_0$ and $f_1$ as a function of momentum, at  time $\tau-\tau_0 =0.5\, \tau_R$. Clearly, even the truncation $N=1$ reproduces accurately the values of these functions over the entire momentum range. As we did in the non expanding case, we can define an effective temperature $T_{\rm eff} $ and an effective chemical potential $\mu_{\rm eff}$ by fitting $f_0(p)$ to a local equilibrium distribution function. The values of the effective parameters thus obtained are plotted in Fig.~\ref{fig:Tmueff_Exp1}. The time-dependence of the effective temperature is compared to that expected for a system in local equilibrium, namely $T(\tau)\sim \tau^{-1/3}$ (ideal hydrodynamical behavior). One sees that this behavior is well reproduced for times $\tau-\tau_0\gtrsim \tau_R$. 

\begin{figure}[H]
 \begin{center}
  \includegraphics[clip,width=8cm]{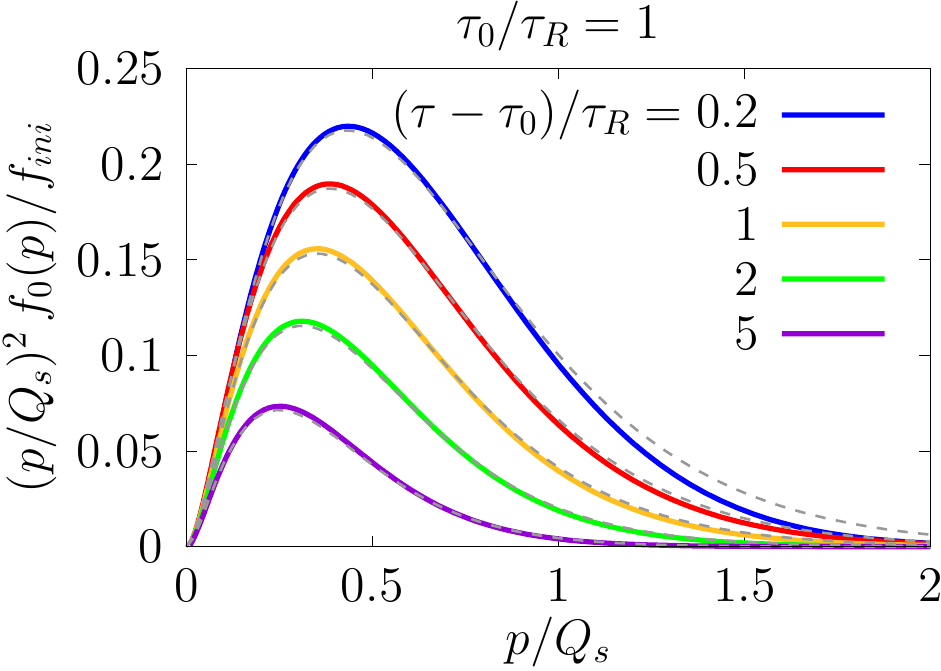}
  \vspace{-20pt}
 \end{center}
 \caption{Time evolution of $f_0 (p)$ multiplied by $p^2$, computed within the truncation  $N=4$. The dashed lines represent the local equilibrium distribution functions with the effective temperature and chemical potential determined as explained in the text. }
 \label{fig:tdep_f0_Exp1}
\end{figure}

  The mode function $f_0(p)$ is  plotted (multiplied by $p^2$) in Fig.~\ref{fig:tdep_f0_Exp1}. This evolution follows a smooth pattern, and in fact is essentially that of an equilibrium distribution function with slowly evolving parameters $T_{\rm eff}$ and $\mu_{\rm eff}$. As can be seen on this figure, by time $\tau-\tau_0\gtrsim \tau_R$, $f_0(p)$ is accurately described by a local equilibrium distribution. However, it is important to note that the mode function $f_1(p)$ does not vanish  at this time, and indeed the longitudinal pressures are not yet equilibrated by this time as can be seen in Fig.~\ref{fig:Ia_exp2}, right panel. 
  
\subsubsection{Smaller collision rates $(\tau_0 /\tau_R =0.1,\ \tau_0 /\tau_R =0.01)$}

We now examine how the system evolves when the collision rate is small compared to the expansion rate. We shall consider two cases. 

\begin{figure}[H]
% \begin{center}
%  \includegraphics[clip,width=7cm]{./fig/Ia_Exp2.pdf}\qquad  \includegraphics[clip,width=7cm]{./fig/Ia_Exp3.pdf}
%     \includegraphics[clip,width=7.5cm]{./fig/energy_Exp2_log.pdf}\qquad  \includegraphics[clip,width=7.5cm]{./fig/energy_Exp3_log.pdf}
%      \includegraphics[clip,width=7cm]{./fig/pressure_ratio_Exp2_log.pdf}\qquad \includegraphics[clip,width=7cm]{./fig/pressure_ratio_Exp3_log.pdf}
%  \vspace{-20pt}
% \end{center}
 \begin{tabular}{cc}
  \begin{minipage}{0.5\hsize}
   \includegraphics[clip,width=7.5cm]{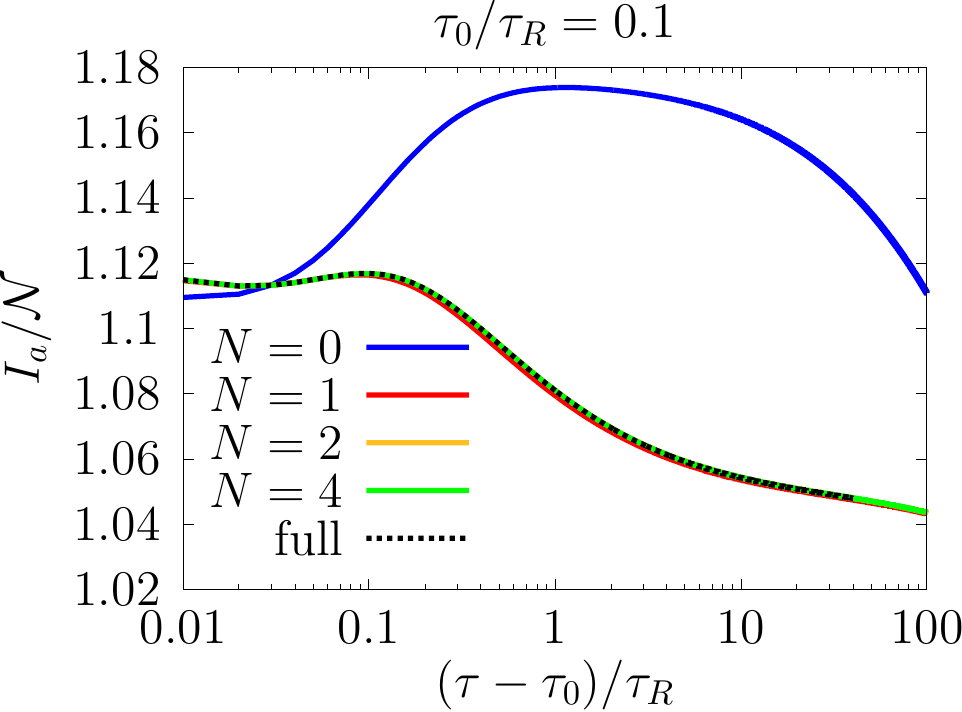} 
  \end{minipage} &
  \begin{minipage}{0.5\hsize}
   \includegraphics[clip,width=7.5cm]{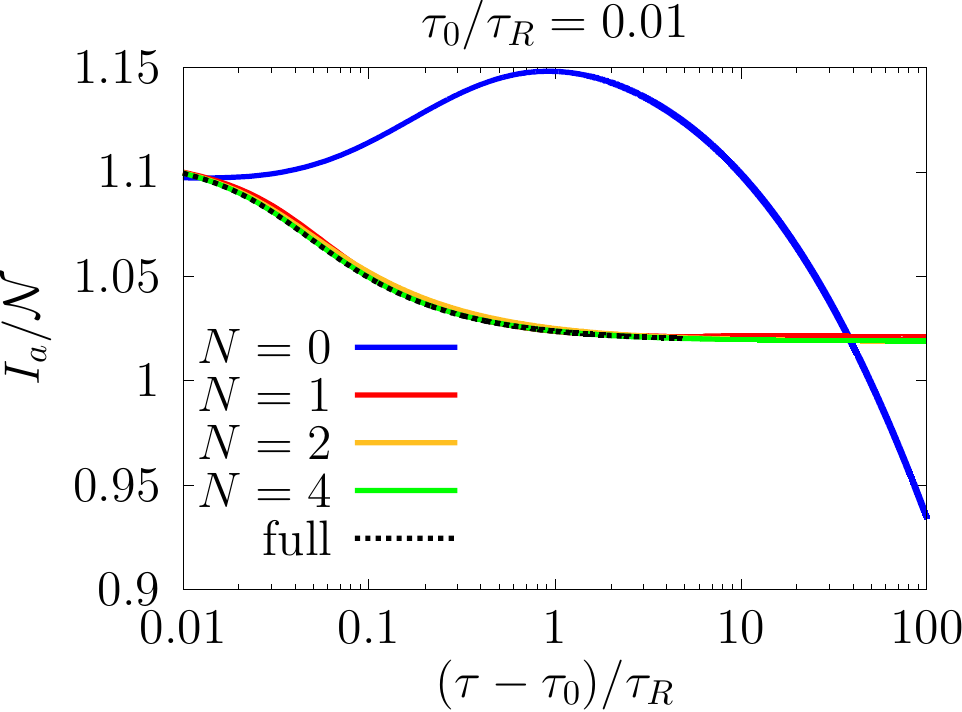}
  \end{minipage} \\
  {} & {} \\[-5pt]
  \begin{minipage}{0.5\hsize}
   \includegraphics[clip,width=7.5cm]{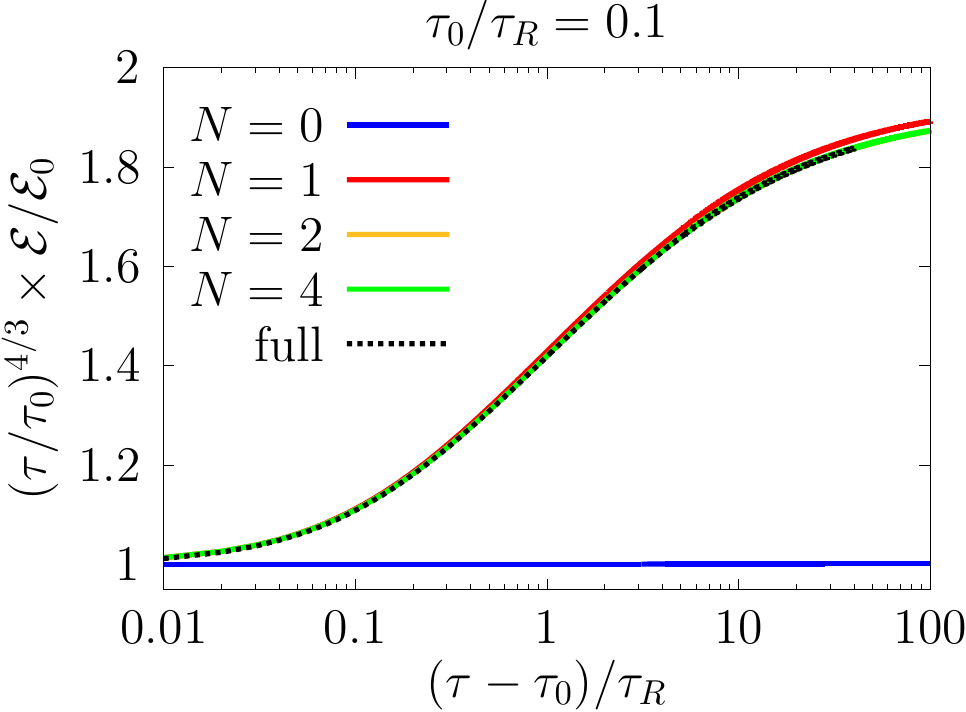}
  \end{minipage} &
  \begin{minipage}{0.5\hsize}
   \includegraphics[clip,width=7.5cm]{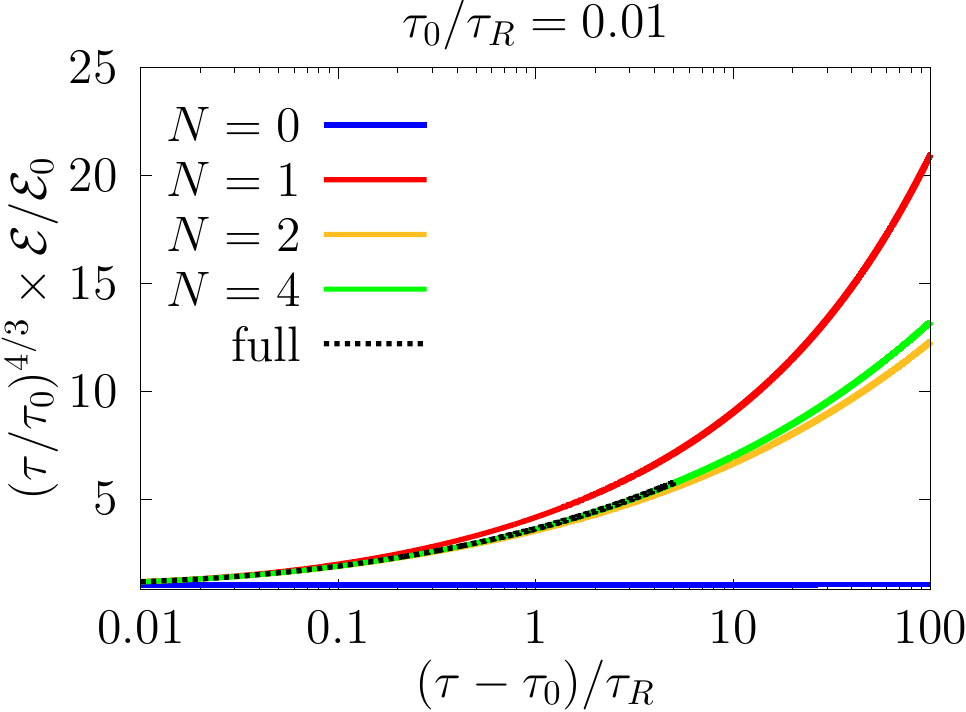}
  \end{minipage} \\
  {} & {} \\[-5pt]
  \begin{minipage}{0.5\hsize}
   \includegraphics[clip,width=7.5cm]{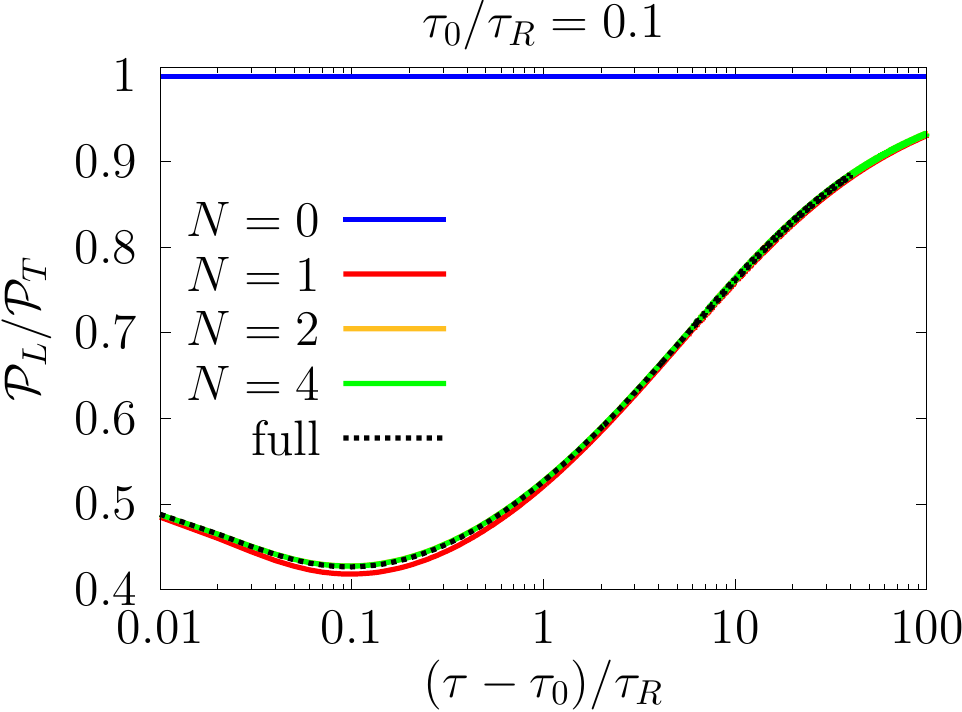}
  \end{minipage} &
  \begin{minipage}{0.5\hsize}
   \includegraphics[clip,width=7.5cm]{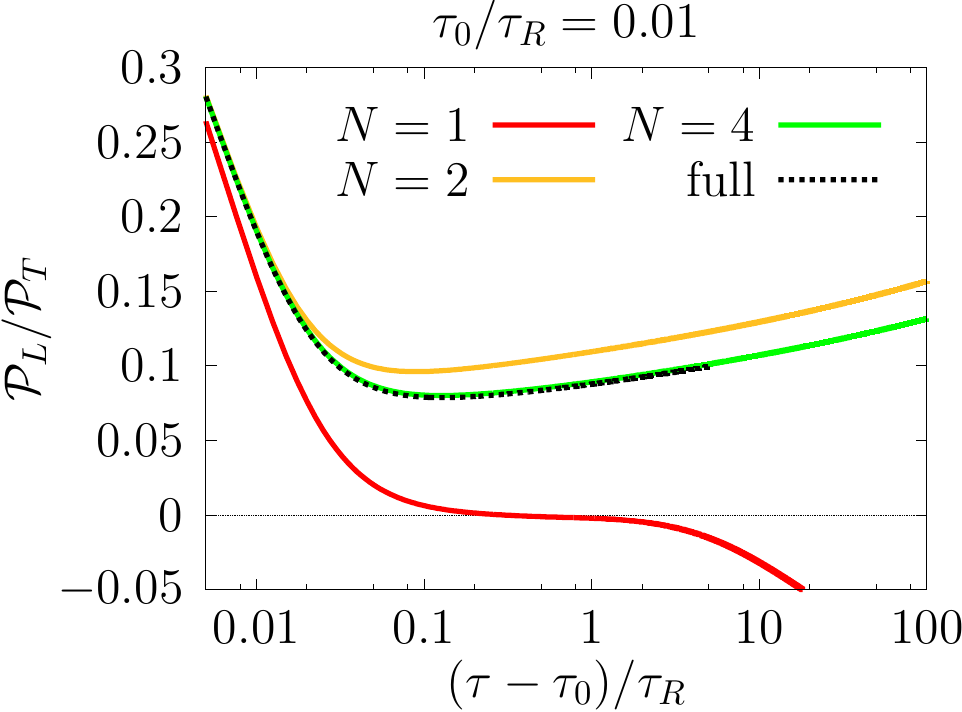}
  \end{minipage}
  \end{tabular}
 \caption{Plot of various quantities in the expanding system, for two different values of the ratio $\tau_0/\tau_R=0.1$ (left) and $\tau_0/\tau_R=0.01$ (right). Several truncation with $N=0,1,2,4$ are compared to the full result (dotted lines).}
 \label{fig:Ia_exp2b}
\end{figure}

\begin{figure}[H]
 \begin{tabular}{cc}
  \begin{minipage}{0.5\hsize}
   \includegraphics[clip,width=7.5cm]{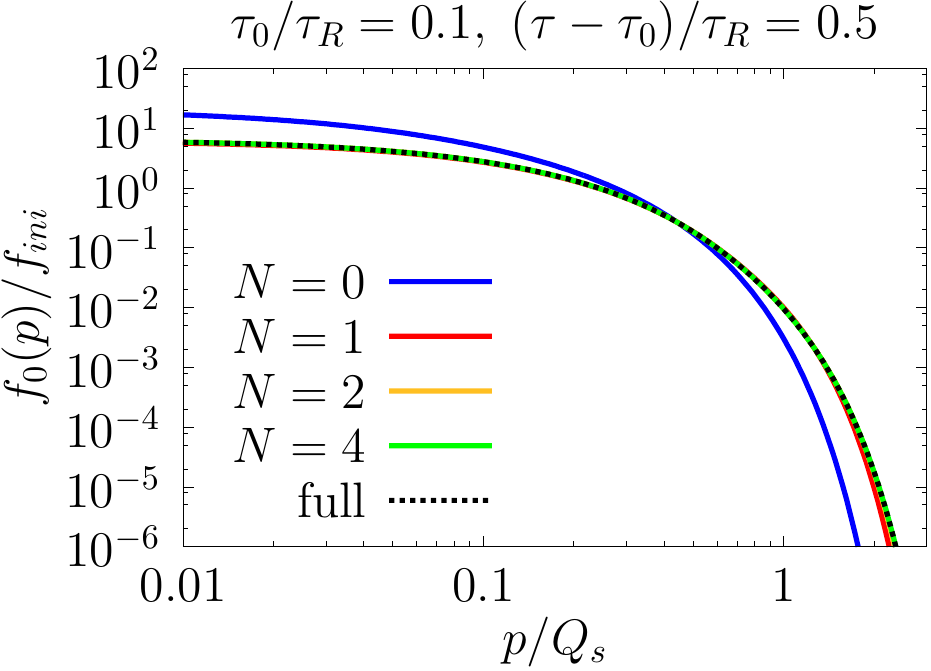} 
  \end{minipage} &
  \begin{minipage}{0.5\hsize}
   \includegraphics[clip,width=7.5cm]{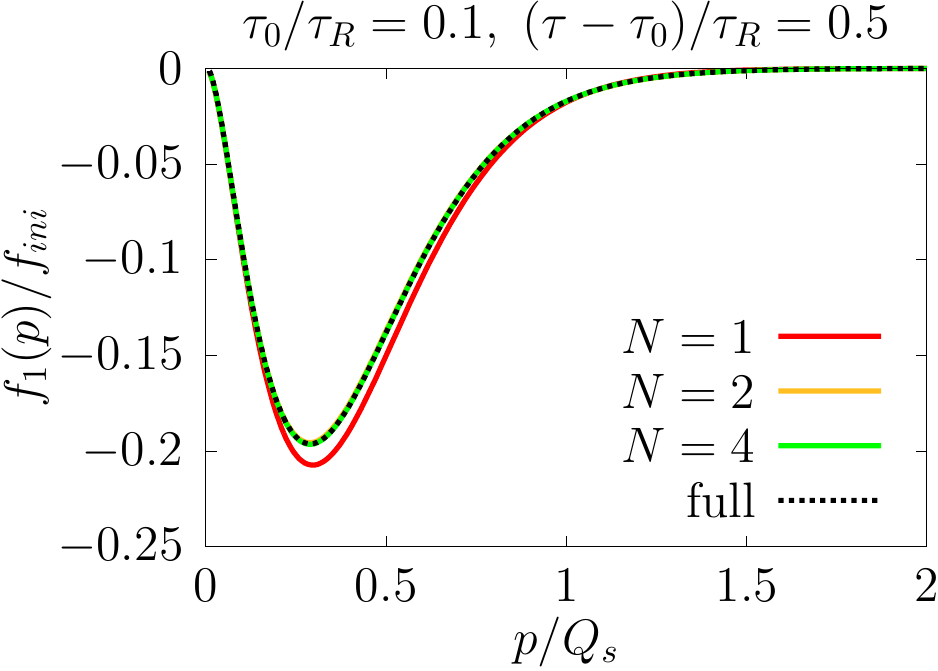}
  \end{minipage} \\
  {} & {} \\[-5pt]
  \begin{minipage}{0.5\hsize}
   \includegraphics[clip,width=7.5cm]{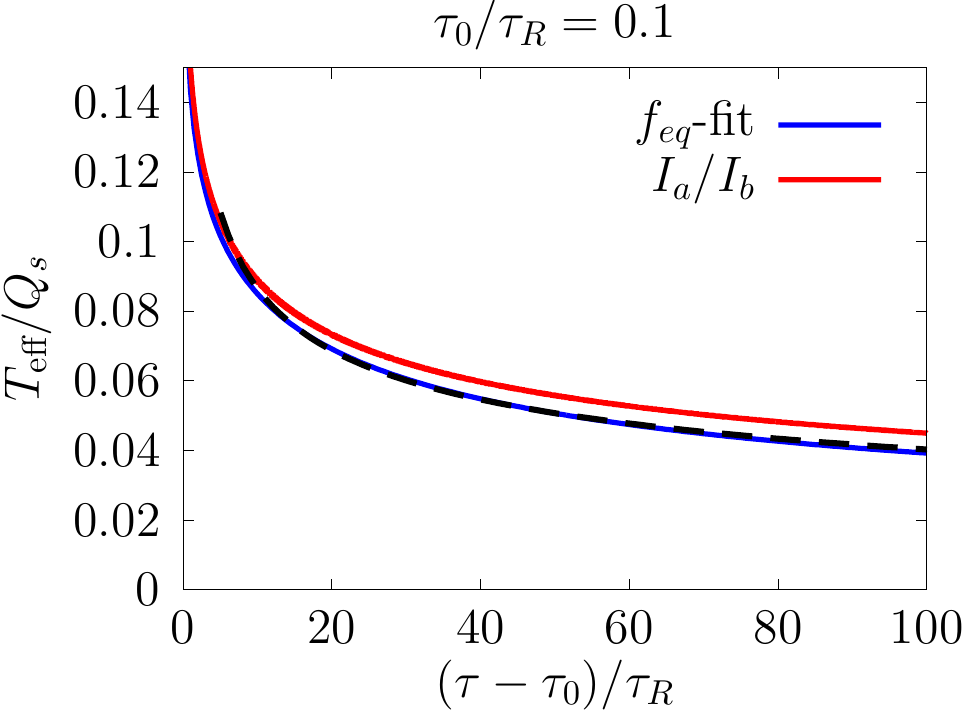}
  \end{minipage} &
  \begin{minipage}{0.5\hsize}
   \includegraphics[clip,width=7.5cm]{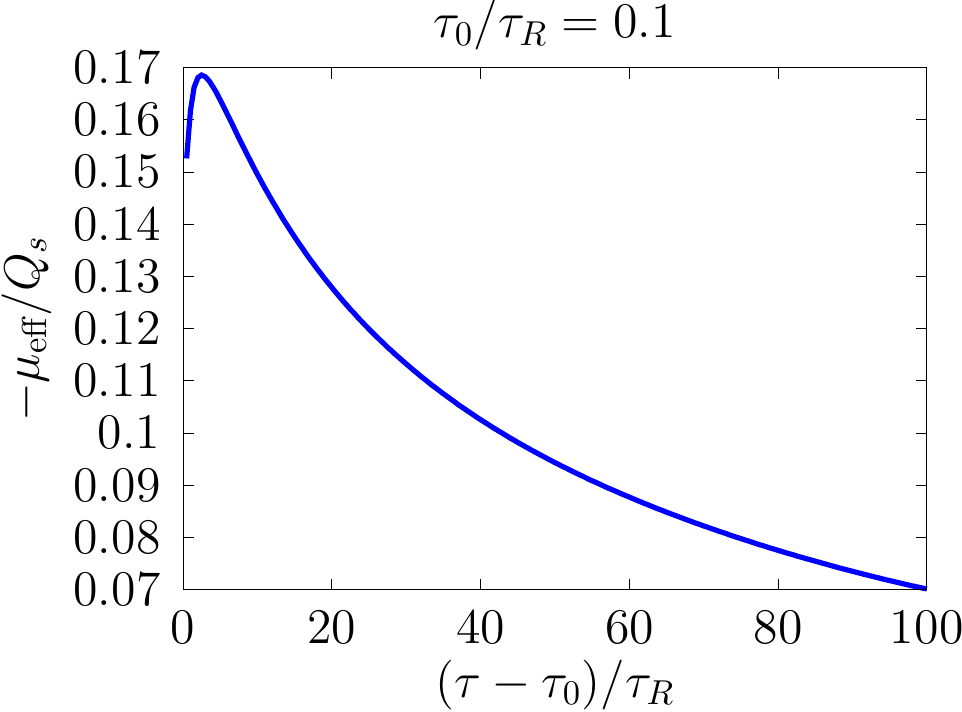}
  \end{minipage}
  \end{tabular}
     \caption{Top: the mode functions $f_0(p)$ (left) and $f_1(p)$ (right) as a function of $p$ in the expanding system, for  $\tau_0/\tau_R=0.1$ and at some early time, for various truncations. The lower panels display the effective temperature and chemical potential as a function of time. The dashed line in the plot of $T_{\rm eff}$ indicates the $\tau^{-1/3}$ dependence of the temperature expected in ideal hydrodynamics.  \label{fig:Ndep_fn_Exp2} }
\end{figure}

\begin{figure}[H]
  \begin{center}
     \includegraphics[clip,width=7.5cm]{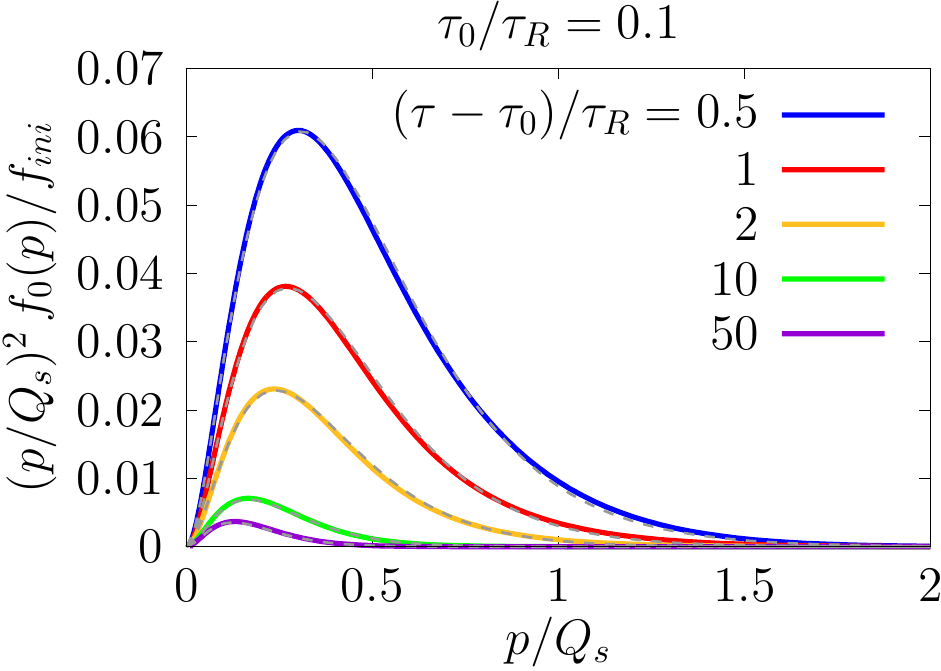}\qquad 
     \includegraphics[clip,width=7.5cm]{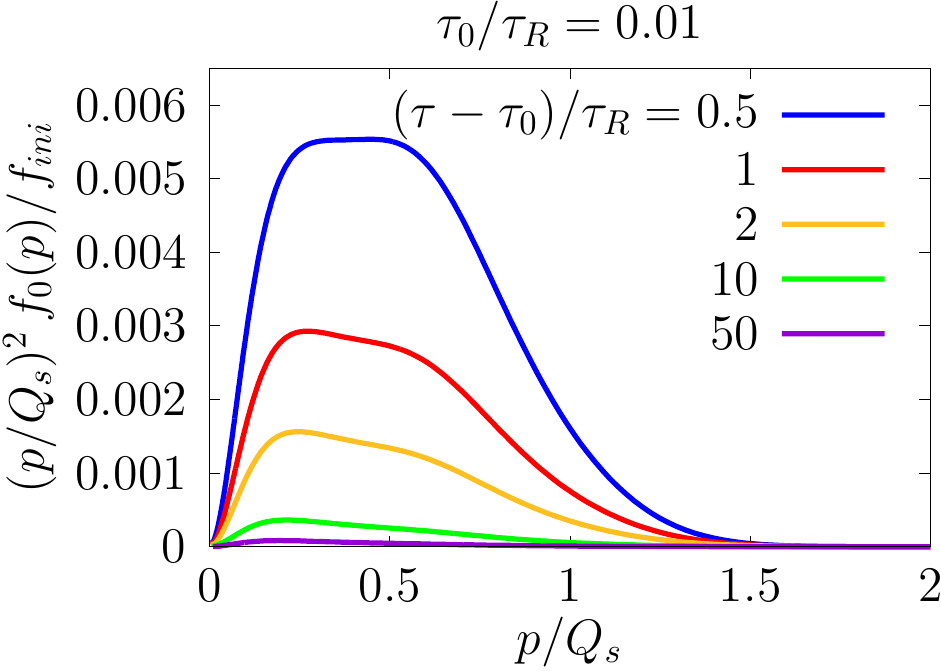}
  \end{center}
 \caption{The mode functions $f_0(p)$ multiplied by $p^2$, as a function of $p$,  for  $\tau_0/\tau_R=0.1$ (left)   and $\tau_0/\tau_R=0.01$ (right). 
The local equilibrium distribution functions with the effective temperature and chemical potential are plotted as dashed lines only on the left panel.
Clearly, even at the latest times, the distribution on the right panel does not look like a thermal distribution.}
 \label{fig:Ndep_fn_Exp2b}
\end{figure}

In the first case, $\tau_0 /\tau_R =0.1$, the results obtained differ only slightly from those of the case $\tau_0 /\tau_R =1$ that we have just examined: the truncations loose accuracy, but even the   truncation   $N=1$ still yields results that compare well with the exact solution. In particular, isotropization and thermalization are clearly achieved at late times. This is illustrated in the left panels of Figs.~\ref{fig:Ia_exp2b} and \ref{fig:Ndep_fn_Exp2b}, which compare well with the corresponding figures in the previous subsection. 
This is also confirmed by the plot of the mode functions $f_0$ and $f_1$ displayed in Fig.~\ref{fig:Ndep_fn_Exp2}, as well as those of the effective  temperature and chemical potential.

\begin{figure}[H]
 \begin{tabular}{cc}
  \begin{minipage}{0.5\hsize}
   \includegraphics[clip,width=7.5cm]{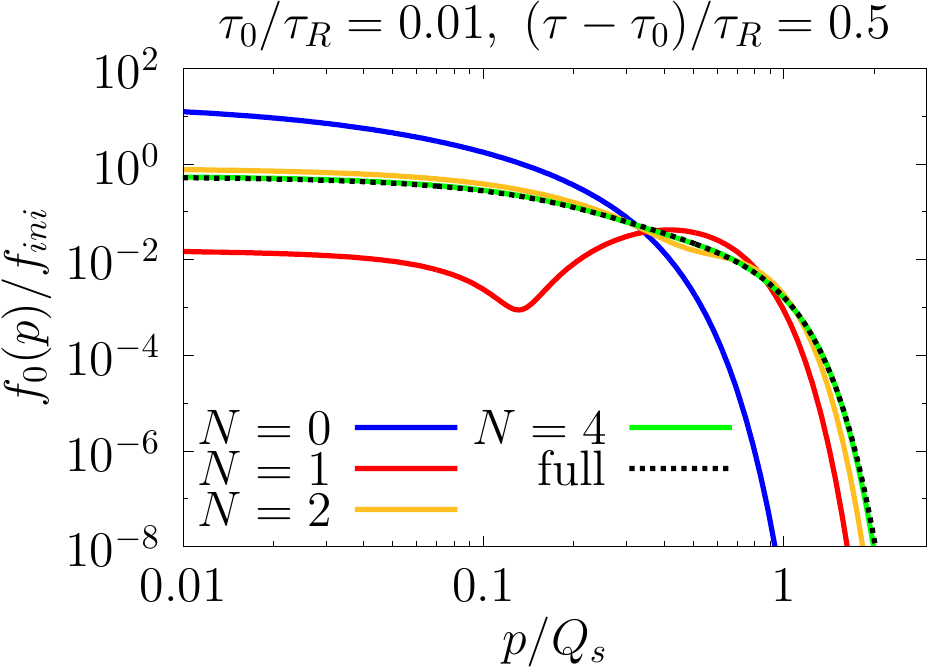} 
  \end{minipage} &
  \begin{minipage}{0.5\hsize}
   \includegraphics[clip,width=7.5cm]{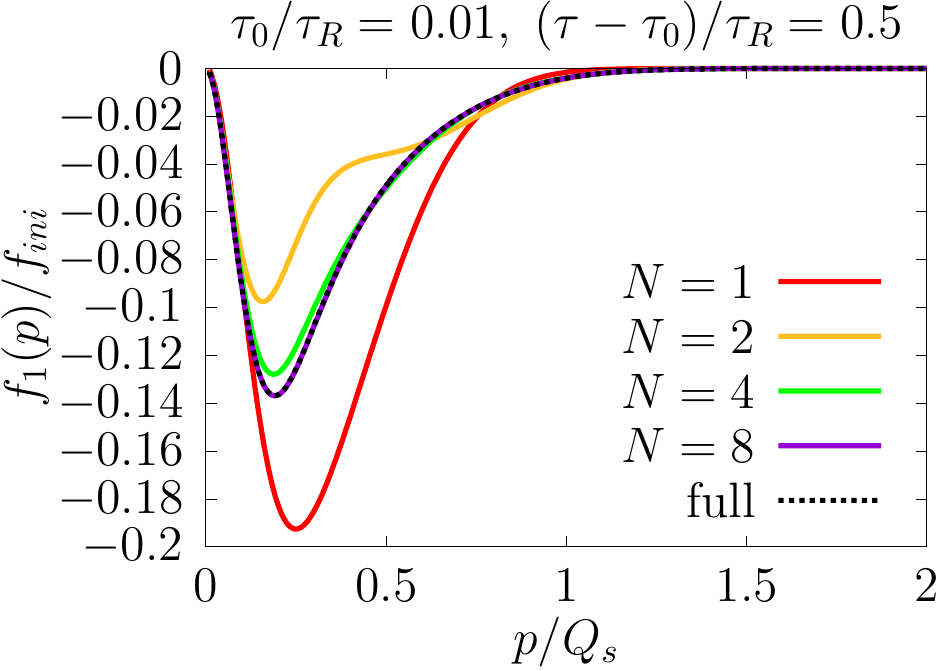}
  \end{minipage} \\
  {} & {} \\[-5pt]
  \begin{minipage}{0.5\hsize}
   \includegraphics[clip,width=7.5cm]{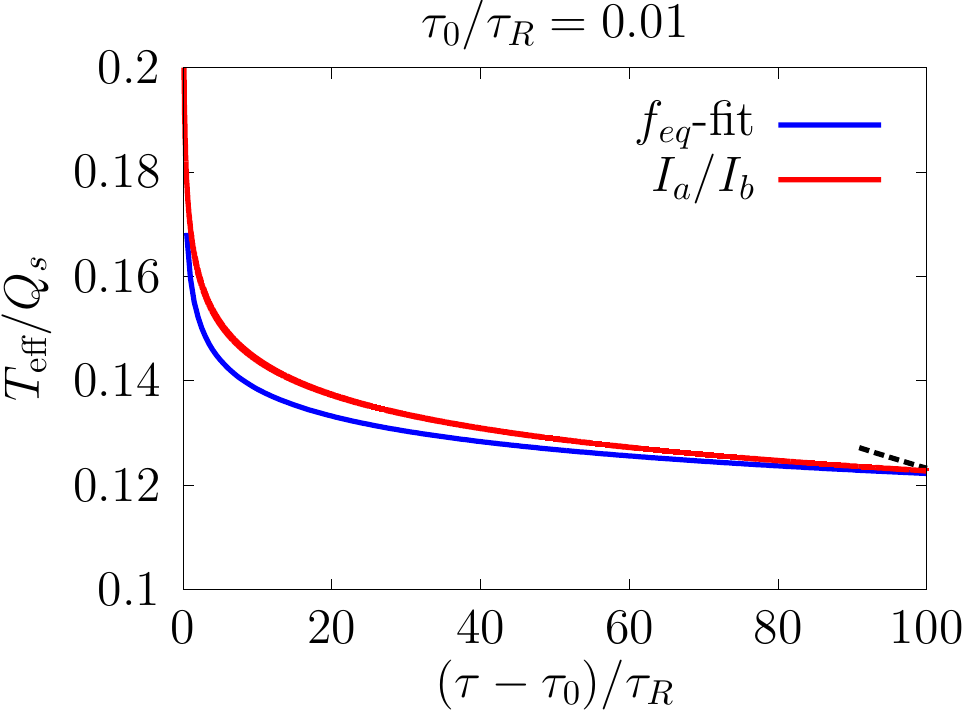}
  \end{minipage} &
  \begin{minipage}{0.5\hsize}
   \includegraphics[clip,width=7.5cm]{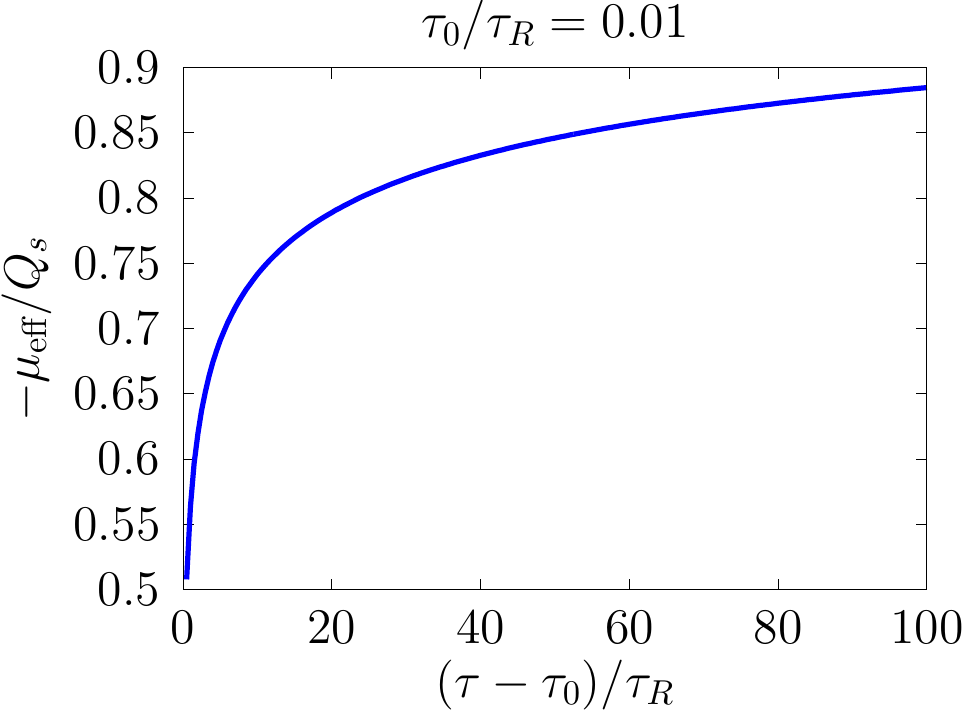}
  \end{minipage}
  \end{tabular}
     \caption{Top: the mode functions $f_0(p)$ (left) and $f_1(p)$ (right) as a function of $p$ in the expanding system, for  $\tau_0/\tau_R=0.01$ and at some early time, for various truncations. The lower panels display the effective temperature and chemical potential as a function of time. The dashed line in the plot of $T_{\rm eff}$ indicates the $\tau^{-1/3}$ dependence of the temperature expected in ideal hydrodynamics, a regime that is clearly not reached in the present case.  \label{fig:Ndep_fn_Exp3} }
\end{figure}

When $\tau_0 /\tau_R =0.01$, the second case that we consider, we enter a regime where more mode functions are needed to reproduce the exact results. This case is illustrated by the plots in the right-hand panels of Figs.~\ref{fig:Ia_exp2b} and \ref{fig:Ndep_fn_Exp2b}. The truncation $N=1$ is no longer sufficient, as indeed some artefacts appear in this case: for instance the longitudinal pressure becomes negative (see the right bottom panel of Fig.~\ref{fig:Ia_exp2b}), an unphysical feature already observed in the two-moment truncation of Ref.~\cite{Blaizot:2019}. Note, however, that from a practical point of view, the expansion in mode functions remains very useful since in all cases only a few modes are needed (in the present case $N=4$ is sufficient to reproduce with  good accuracy all the quantities considered, see e.g. Fig.~\ref{fig:Ndep_fn_Exp3}. We note also that in this case of small collision rate, the thermalization of the system is not achieved during the time span of the calculation. In particular the mode function $f_0(p)$ is never well approximated by a local equilibrium function as can be seen in the right panel of Fig.~\ref{fig:Ndep_fn_Exp2b}. It is also clear from the plot of $T_{\rm eff}$ in Fig.~\ref{fig:Ndep_fn_Exp3} that the hydrodynamic regime is not reached.

%%%%%%%%%%%%%%%%%%%%%%%%%%%%%%%%%%%%%%%%%%%%%%%%%%%%%%%%%%%%%%%%%%%%%%%%%%%%%%%%%%%%%%%
\section{Conclusions}

We have shown that the expansion of the distribution function in angular modes allows for a significant reduction of the numerical effort involved in solving the Boltzmann equation for longitudinally expanding quark-gluon plasmas. This is because only a few angular modes are enough to capture the main features of the time evolution of the energy density and the pressures, which are related to the lowest two moments of the distribution function, i.e. to the first two angular modes $f_0$ and $f_1$. The number of modes needed for an accurate description of the distribution function depends on the ratio between the expansion rate and the collision rate. For most cases of physical interest, the first two modes suffice, higher modes being needed only in cases of small collision rates, i.e., when the evolution of the system is strongly dominated by free streaming. 

We have presented results of numerical calculations, first for a non expanding system, and then for a boost invariant, longitudinally expanding system. In both cases we found a rapid convergence of the expansion. In the non-expanding case, we observe the isotropization of the distribution, measured by the decay of the various mode functions $f_{n\ge 1}$. The thermalization is observed though the fact  that the spherical mode $f_0(p)$  is well described at late time  by a thermal distribution, with slowly evolving temperature and chemical potential.  In the expanding system, the expansion delays the effect of the collisions. We find however that this delay alters the qualitative pattern only when the collision rate is much smaller than the expansion rate. When both rates are comparable, we find that the system approaches equilibrium in very much the same way as in the non-expanding case.  However, the pressure anisotropy takes more time to disappear, in particular it subsists  even after the spherical mode is well described by a thermal distribution. This is due to the fact that $f_0$ and $f_1$ remain strongly coupled, and the expansion hinders the decay of $f_1$. When the collision rate is too small, the thermalization itself is hindered by the fast expansion. 

These conclusions strengthen those obtained in \cite{Blaizot:2017ucy,Blaizot:2019}, based on specific moments of the distribution function and a relaxation time approximation for the collision kernel.  
The present study should be extended to more realistic collision kernel involving both elastic and inelastic processes. This is work for the future. \\

{\noindent \bf Acknowledgments}

We thank Li Yan for useful exchanges at an early stage of this project. We also thank Fran\c cois Gelis for comments on the manuscript.

\appendix
%%%%%%%%%%%%%%%%%%%%%%%%%%%%%%%%%%%%%%%%%%%%%%%%%%%%%%%%%%%%%%%%%%%%%%%%%%%%%%%%%%%%%%%
\section{Equations for the moments} \label{sec:moment}
In this Appendix, we derive the equation of motion for the moments $\mathcal{L}_n$ defined in Eq.~\eqref{def_momL} from the Boltzmann equation \eqref{Boltz0} with the small-angle approximation for the collision term \eqref{col_sa0}.

From the Boltzmann equation \eqref{Boltz0}, it follows that
\begin{align}
\frac{\partial}{\partial \tau} \mathcal{L}_n 
= \frac{1}{\tau} \int\! \frac{d^3 p}{(2\pi)^3} \, p P_{2n} (p_z /p ) p_z \frac{\partial}{\partial p_z} f(\bp )
-\int\! \frac{d^3 p}{(2\pi)^3} \, p P_{2n} (p_z /p ) C[f(\bp )] \, .
\end{align}
By applying integration by parts and the recursion relations for the Legendre polynomials
\begin{equation}
x P_n (x) = \frac{n+1}{2n+1} P_{n+1} (x) +\frac{n}{2n+1} P_{n-1} (x) \label{LP_rec1}
\end{equation} 
and
\begin{equation}
\frac{1-x^2}{n} \frac{d}{dx} P_n (x) = -x P_n (x) +P_{n-1} (x) \, , \label{LP_rec2}
\end{equation}
one can compute the moments of the drift term as \cite{Blaizot:2017ucy}
\begin{equation}
\frac{1}{\tau} \int\! \frac{d^3 p}{(2\pi)^3} \, p P_{2n} (p_z /p ) p_z \frac{\partial}{\partial p_z} f(\bp )
= -\frac{1}{\tau} \left( \alpha_n \mathcal{L}_n +\beta_n \mathcal{L}_{n-1} +\gamma_n \mathcal{L}_{n+1} \right) \, ,
\end{equation}
with coefficients
\begin{align}
\alpha_n &= \frac{2(14n^2+7n-2)}{(4n-1)(4n+3)} \\
\beta_n &= \frac{(2n-1)2n(2n+2)}{(4n-1)(4n+1)} \\
\gamma_n &= -\frac{(2n-1)(2n+1)(2n+2)}{(4n+1)(4n+3)} \, . 
\end{align}
These coefficients are related with those given by Eqs.~(\ref{coeff_first}-\ref{coeff_last}) as $\alpha_n = 4a_n -\tilde{a}_n$, $\beta_n = 4b_n -\tilde{b}_n$ and $\gamma_n = 4c_n -\tilde{c}_n $. 

The moment of the collision term can be computed in a similar way. We find
\begin{align}
&\int\! \frac{d^3 p}{(2\pi)^3} \, p P_{2n} (p_z /p ) C[f(\bp )] \notag \\
&= 2(2n^2 +n-1) I_a \int\! \frac{d^3 p}{(2\pi)^3} \, \frac{1}{p} P_{2n} (p_z /p) f(\bp ) +I_b \int\! \frac{d^3 p}{(2\pi)^3} \, P_{2n} (p_z /p) f(\bp) \left[ 1+f(\bp)\right] \, .
\end{align}
The momentum integrals that appear in the right hand side are different kinds of moments from $\mathcal{L}_n$. For those, let us introduce 
\begin{equation}
\mathcal{J}_n = \int\! \frac{d^3 p}{(2\pi)^3} \, \frac{1}{p} P_{2n} (p_z /p) f(\bp ) \, ,
\end{equation}
and
\begin{equation}
\mathcal{K}_n = \int\! \frac{d^3 p}{(2\pi)^3} \, P_{2n} (p_z /p) f(\bp) \left[ 1+f(\bp)\right] \, .
\end{equation}
Finally, the equation for the moments $\mathcal{L}_n$ reads
\begin{equation}
\frac{\partial}{\partial \tilde{\tau}} \mathcal{L}_n 
= -\frac{1}{\tilde{\tau}} \left( \alpha_n \mathcal{L}_n +\beta_n \mathcal{L}_{n-1} +\gamma_n \mathcal{L}_{n+1} \right) 
-\left[ 2(2n^2 +n-1) \mathcal{K}_0 \mathcal{J}_n +2\mathcal{J}_0 \mathcal{K}_n \right] \, .
\end{equation}
This equation is clearly not closed within the moments $\mathcal{L}_n$. If one tries deriving similar equations for the moments $\mathcal{J}_n$ and $\mathcal{K}_n$, one encounters new kinds of moments. This makes it difficult to compute the dynamics of this system based on simple equations for a closed set of  moments.

%%%%%%%%%%%%%%%%%%%%%%%%%%%%%%%%%%%%%%%%%%%%%%%%%%%%%%%%%%%%%%%%%%%%%%%%%%%%%%%%%%%%%%%
\section{Derivation of the kinetic equations for the angular mode functions} \label{sec:derivation}
By applying  the mode projection operator $\frac{4n+1}{2} \int_{-1}^1 d(\cos \theta) \, P_{2n} (\cos \theta)$ to the Boltzmann equation \eqref{Boltz0}, we can derive the equations of motion for the mode function $f_n (p)$. 
First, we rewrite the Boltzmann equation in terms of the variables $p=|\bp|$ and $x=\cos \theta$,
\begin{equation}
\frac{\partial}{\partial \tau} f(p,x) 
= \frac{1}{\tau} \left[ px^2 \frac{\partial}{\partial p} +x(1-x^2) \frac{\partial}{\partial x} \right] f(p,x) -C[f] \, ,
\end{equation}
with
\begin{align}
C[f(p,x)] 
&= -I_a \frac{1}{p^2} \left[\frac{\partial}{\partial p} p^2 \frac{\partial}{\partial p}
+\frac{\partial}{\partial x} (1-x^2) \frac{\partial}{\partial x} \right] f(p,x) 
-I_b \frac{1}{p^2} \frac{\partial}{\partial p} p^2 f(p,x) (1+f(p,x)) \, .
\end{align}

From Eqs.~\eqref{LP_rec1}, \eqref{LP_rec2} and the orthogonality relation
\begin{equation}
\int_{-1}^1 P_n (x) P_m (x) dx = \frac{2}{2n+1} \delta_{n,m} \, , \label{LP_ortho}
\end{equation}
one can derive the following formulas:
\begin{align}
&\frac{4n+1}{2} \int_{-1}^1 dx \, P_{2n} (x) x^2 P_{2m} (x) \notag \\
&= \frac{(2n+1)(2n+2)}{(4n+3)(4n+5)} \delta_{n+1,m} +\frac{8n^2+4n-1}{(4n-1)(4n+3)} \delta_{n,m} +\frac{(2n-1)2n}{(4n-3)(4n-1)} \delta_{n-1,m} \, ,
\end{align}
and
\begin{align}
&\frac{4n+1}{2} \int_{-1}^1 dx \, P_{2n} (x) x(1-x^2) \frac{\partial}{\partial x} P_{2m} (x) \notag \\
&= -\frac{(2n-2)(2n-1)2n}{(4n-3)(4n-1)} \delta_{n,m+1} +\frac{2n(2n+1)}{(4n-1)(4n+3)} \delta_{n,m} 
+\frac{(2n+1)(2n+2)(2n+3)}{(4n+3)(4n+5)} \delta_{n,m-1} \, .
\end{align}
By using these formulas, one can find the projection of the expansion drift term as is given in Eq.~\eqref{Boltz_n}. 

Next, we compute the projection of the collision term. The term proportional to $I_a$ is simple because the Legendre polynomials are eigen functions of the Laplacian.
\begin{align}
\frac{4n+1}{2} \int_{-1}^1 dx \, P_{2n} (x) \frac{1}{p^2} \left[\frac{\partial}{\partial p} p^2 \frac{\partial}{\partial p}
+\frac{\partial}{\partial x} (1-x^2) \frac{\partial}{\partial x} \right] f(p,x) 
= \frac{1}{p^2} \left[\frac{\partial}{\partial p} p^2 \frac{\partial}{\partial p}
-2n(2n+1) \right] f_n (p) \, .
\end{align}
The term proportional to $I_b$ is much more involved since it is nonlinear in $f$.
\begin{align}
&\frac{4n+1}{2} \int_{-1}^1 dx \, P_{2n} (x) \frac{1}{p^2} \frac{\partial}{\partial p} p^2 f(p,x) (1+f(p,x)) \notag \\
&\hspace{20pt}
= \frac{1}{p^2} \frac{\partial}{\partial p} p^2 \left[ f_n (p)
+\frac{4n+1}{2} \sum_{m,l=0}^\infty A_{n,m,l} f_m(p) f_l(p) \right] \, ,
\end{align}
where we have defined
\begin{equation}
A_{n,m,l} = \int_{-1}^1 dx \, P_{2n} (x) P_{2m} (x) P_{2l} (x) \, .
\end{equation}
The explicit form of $A_{n,m,l}$ can be derived by using the formula [Eq.~8.915.5 (p.986) of Ref.~\cite{gradshteyn2007}]
\begin{equation}
P_n (x) P_m (x)
= \sum_{k=0}^{m} \frac{\alpha_{m-k} \alpha_k \alpha_{n-k}}{\alpha_{n+m-k}} \frac{2n+2m-4k+1}{2n+2m-2k+1} P_{n+m-2k} (x)
\hspace{20pt} [m\leq n] \, ,
\label{PPexpansion}
\end{equation}
where
\begin{equation}
\alpha_k = \frac{(2k-1)!!}{k!} \, . 
\end{equation}
Here, $\alpha_0$ should be understood as 1.
Thanks to this formula, we can show
\begin{align}
A_{n,m,l}
&= \frac{2}{4n+1} \sum_{k=0}^{\text{min}(2m,2l)} 
\frac{\alpha_{2l-k} \alpha_k \alpha_{2m-k}}{\alpha_{2m+2l-k}} \frac{4m+4l-4k+1}{4m+4l-2k+1} \delta_{n, m+l-k} \notag \\
&= \begin{cases}
 \displaystyle
 \frac{2}{2n+2m+2l+1} \frac{\alpha_{n+m-l} \alpha_{n-m+l} \alpha_{-n+m+l}}{\alpha_{n+m+l}} 
 & \text{if} \ n+m-l, \, n-m+l, \, -n+m+l \geq 0  \\
0 & \text{otherwise}.
\end{cases} 
\end{align}
Collecting all the above results, we arrive in Eq.~\eqref{Boltz_n}.

%%%%%%%%%%%%%%%%%%%%%%%%%%%%%%%%%%%%%%%%%%%%%%%%%%%%%%%%%%%%%%%%%%%%%%%%%%%%%%%%%%%%%%%
%% bibliography
\bibliographystyle{h-physrev3} 
\bibliography{ref}

\begin{thebibliography}{10}

\bibitem{Xu:2004mz}
Z.~Xu and C.~Greiner,
\newblock Phys. Rev. {\bf C71}, 064901 (2005), hep-ph/0406278.
%%CITATION = HEP-PH/0406278;%%

\bibitem{Scardina:2014gxa}
F.~Scardina, D.~Perricone, S.~Plumari, M.~Ruggieri, and V.~Greco,
\newblock Phys. Rev. {\bf C90}, 054904 (2014), 1408.1313.
%%CITATION = ARXIV:1408.1313;%%

\bibitem{Kurkela:2015qoa}
A.~Kurkela and Y.~Zhu,
\newblock Phys. Rev. Lett. {\bf 115}, 182301 (2015), 1506.06647.
%%CITATION = ARXIV:1506.06647;%%

\bibitem{Mazeliauskas:2018yef}
A.~Mazeliauskas and J.~Berges,
\newblock Phys. Rev. Lett. {\bf 122}, 122301 (2019), 1810.10554.
%%CITATION = ARXIV:1810.10554;%%

\bibitem{Bjorken:1982qr}
J.~D. Bjorken,
\newblock Phys. Rev. {\bf D27}, 140 (1983).
%%CITATION = PHRVA,D27,140;%%

\bibitem{Blaizot:2017ucy}
J.-P. Blaizot and L.~Yan,
\newblock Phys. Lett. {\bf B780}, 283 (2018), 1712.03856.
%%CITATION = ARXIV:1712.03856;%%

\bibitem{Blaizot:2019}
J.-P. Blaizot and L.~Yan,
\newblock in preparation .

\bibitem{Blaizot:2017lht}
J.-P. Blaizot and L.~Yan,
\newblock JHEP {\bf 11}, 161 (2017), 1703.10694.
%%CITATION = ARXIV:1703.10694;%%

\bibitem{Mueller:1999pi}
A.~H. Mueller,
\newblock Phys. Lett. {\bf B475}, 220 (2000), hep-ph/9909388.
%%CITATION = HEP-PH/9909388;%%

\bibitem{Bjoraker:2000cf}
J.~Bjoraker and R.~Venugopalan,
\newblock Phys. Rev. {\bf C63}, 024609 (2001), hep-ph/0008294.
%%CITATION = HEP-PH/0008294;%%

\bibitem{Blaizot:2013lga}
J.-P. Blaizot, J.~Liao, and L.~McLerran,
\newblock Nucl. Phys. {\bf A920}, 58 (2013), 1305.2119.
%%CITATION = ARXIV:1305.2119;%%

\bibitem{Denicol:2012cn}
G.~S. Denicol, H.~Niemi, E.~Molnar, and D.~H. Rischke,
\newblock Phys. Rev. {\bf D85}, 114047 (2012), 1202.4551,
\newblock [Erratum: Phys. Rev.D91,no.3,039902(2015)].
%%CITATION = ARXIV:1202.4551;%%

\bibitem{Behtash:2017wqg}
A.~Behtash, C.~N. Cruz-Camacho, and M.~Martinez,
\newblock Phys. Rev. {\bf D97}, 044041 (2018), 1711.01745.
%%CITATION = ARXIV:1711.01745;%%

\bibitem{Tanji:2017suk}
N.~Tanji and R.~Venugopalan,
\newblock Phys. Rev. {\bf D95}, 094009 (2017), 1703.01372.
%%CITATION = ARXIV:1703.01372;%%

\bibitem{Epelbaum:2015vxa}
T.~Epelbaum, F.~Gelis, S.~Jeon, G.~Moore, and B.~Wu,
\newblock JHEP {\bf 09}, 117 (2015), 1506.05580.
%%CITATION = ARXIV:1506.05580;%%

\bibitem{Keegan:2015avk}
L.~Keegan, A.~Kurkela, P.~Romatschke, W.~van~der Schee, and Y.~Zhu,
\newblock JHEP {\bf 04}, 031 (2016), 1512.05347.
%%CITATION = ARXIV:1512.05347;%%

\bibitem{Mueller:1999fp}
A.~H. Mueller,
\newblock Nucl. Phys. {\bf B572}, 227 (2000), hep-ph/9906322.
%%CITATION = HEP-PH/9906322;%%

\bibitem{gradshteyn2007}
I.~S. Gradshteyn and I.~M. Ryzhik,
\newblock {\em Table of integrals, series, and products} (Academic press,
  2007).

\end{thebibliography}

\end{document}